\documentclass[pdftex,twocolumn,epjc3]{svjour3}          % twocolumn
\def\makeheadbox{\relax} %added to remove frame on the first page with "will be inserted by editor"
\RequirePackage[T1]{fontenc}
\RequirePackage{fix-cm}

\usepackage[english]{babel}

\smartqed  % flush right qed marks, e.g. at end of proof

\RequirePackage{graphicx}
\RequirePackage{mathptmx}
\usepackage{comment}

\DeclareMathAlphabet{\mathcal}{OMS}{cmsy}{m}{n}

\usepackage{footmisc}
\RequirePackage[numbers,sort&compress]{natbib}
\usepackage{url}
\RequirePackage[colorlinks,citecolor=blue,urlcolor=blue,linkcolor=blue]{hyperref}

\RequirePackage{siunitx}
\RequirePackage{amsmath}
\usepackage{caption}
\usepackage{upgreek}
\usepackage{subfig}
\usepackage{caption}
\usepackage{float}
\usepackage{ragged2e}
\usepackage{pdfpages}
\usepackage{bbold}

\usepackage{bm}

\graphicspath{{./Images/}}

\journalname{Eur. Phys. J. C}

\begin{document}

\raggedbottom

%\title{An active magnetic shielding house for high-precision experiments}
\title{A large `Active Magnetic Shield' for a high-precision experiment}

%\subtitle{nEDM collaboration}
\author{
    C.~Abel\thanksref{Sussex}
    \and
    N.~J.~Ayres\thanksref{ETH}
    \and
    G.~Ban\thanksref{CAEN}  
    \and
    G.~Bison\thanksref{PSI}
    \and
    K.~Bodek\thanksref{Cracow}
    \and
    V.~Bondar\thanksref{ETH,e8} 
    \and        
    T.~Bouillaud\thanksref{LPSC}
    \and
    E.~Chanel\thanksref{Bern,e7}
    \and
    J.~Chen\thanksref{CAEN}
    \and
    W.~Chen\thanksref{ETH,PSI}
    \and
    P.-J. Chiu\thanksref{ETH,PSI,e6}
    \and
    C.~B.~Crawford\thanksref{Kentucky} 
    \and
    M.~Daum\thanksref{PSI}
    \and
    C.~B.~Doorenbos\thanksref{ETH,PSI}
    \and
    S.~Emmenegger\thanksref{ETH,e5}
    \and
    L.~Ferraris-Bouchez\thanksref{LPSC}
    \and
    M.~Fertl\thanksref{Mainz}
    \and
     A.~Fratangelo\thanksref{Bern} 
    \and
    W.~C.~Griffith\thanksref{Sussex} 
    \and
    Z.~D.~Grujic\thanksref{Serbia} 
   \and
    P.~Harris\thanksref{Sussex}
   \and
    K.~Kirch\thanksref{ETH,PSI,e9}
    \and
    V.~Kletzl\thanksref{ETH,PSI}
    \and
    P.~A.~Koss\thanksref{Leuven,e4}
    \and
    J.~Krempel\thanksref{ETH,e10}
    \and
    B.~Lauss\thanksref{PSI}
    \and
    T.~Lefort\thanksref{CAEN}   
    \and
    P.~Mullan\thanksref{ETH}
    \and
    O.~Naviliat-Cuncic\thanksref{CAEN}  
    \and
    D.~Pais\thanksref{ETH,PSI}
    \and
    F.~M.~Piegsa\thanksref{Bern}
    \and
    G.~Pignol\thanksref{LPSC}   
    \and
    M.~Rawlik\thanksref{ETH,e3}
    \and
    I.~Rienäcker\thanksref{PSI}
    \and  
    D.~Ries\thanksref{PSI}
    \and 
    S.~Roccia\thanksref{LPSC}
    \and
    D.~Rozpedzik\thanksref{Cracow}
    \and
    W.~Saenz-Arevalo\thanksref{CAEN}
    \and
    P.~Schmidt-Wellenburg\thanksref{PSI} 
    \and
    A.~Schnabel\thanksref{PTB} 
    \and
    E.~P.~Segarra\thanksref{PSI}
    \and
    N.~Severijns\thanksref{Leuven}
    \and
    T.~Shelton\thanksref{Kentucky} 
    \and
     K.~Svirina\thanksref{LPSC}    
    \and
    R.~Tavakoli Dinani\thanksref{Leuven}
    \and
    J.~Thorne\thanksref{Bern}
    \and
    R.~Virot\thanksref{LPSC}
    \and
    N.~Yazdandoost\thanksref{Mainz2} 
    \and
    J.~Zejma\thanksref{Cracow} 
    \and
    N.~Ziehl\thanksref{ETH}
    \and
    G.~Zsigmond\thanksref{PSI}
    }

\subtitle{nEDM collaboration}

\institute{
    University of Sussex, Department of Physics and Astronomy,Falmer, Brighton BN1 9QH, UK \label{Sussex}
    \and
    ETH Zürich, Institute for Particle Physics and Astrophysics, CH-8093 Zürich, Switzerland \label{ETH}
    \and
    Normandie Univ, ENSICAEN, UNICAEN, CNRS/IN2P3, LPC Caen, 14000 Caen, France \label{CAEN}
    \and
    Paul Scherrer Institut, CH-5232 Villigen PSI, Switzerland \label{PSI}
    \and
    Marian Smoluchowski Institute of Physics, Jagiellonian University, 30-348 Cracow, Poland \label{Cracow}
    \and
    LPSC, Université Grenoble Alpes, CNRS/IN2P3, Grenoble, France\label{LPSC}
    \and
    University of Bern, Albert Einstein Center for Fundamental Physics, CH-3012 Bern, Switzerland\label{Bern}
    \and
    University of Kentucky, Lexington, USA\label{Kentucky}
    \and
    Institute of Physics, Johannes Gutenberg University, D-55128 Mainz, Germany\label{Mainz}
    \and
    Institute of Physics Belgrade, University of Belgrade, 11080 Belgrade, Serbia \label{Serbia}
    \and
    Physikalisch Technische Bundesanstalt, Berlin, Germany\label{PTB}
    \and
    Institute for Nuclear and Radiation Physics, KU Leuven, B-3001 Leuven, Belgium\label{Leuven}
    \and
    Department of Chemistry - TRIGA site, Johannes Gutenberg University, 55128 Mainz, Germany\label{Mainz2}
}
\thankstext{e8}{Corresponding author: 	klaus.kirch@psi.ch}
\thankstext{e9}{Corresponding author: bondarv@phys.ethz.ch}
\thankstext{e10}{Corresponding author: 	 jochen.krempel@phys.ethz.ch}
\thankstext{e7}{Present address: Institut Laue Langevin, 71 avenue des Martyrs CS 20156, 38042 GRENOBLE Cedex 9, France}
\thankstext{e6}{Present address: University of Zurich,
8057 Zurich, Switzerland}
\thankstext{e5}{Present address: Hochschule Luzern, 6002 Luzern, Switzerland}
\thankstext{e4}{Present address: Fraunhofer Institute for Physical Measurement Techniques, 79110 Freiburg, Germany}
\thankstext{e3}{Present address: Paul Scherrer Institut, CH-5232 Villigen PSI, Switzerland}

\maketitle

\begin{abstract}
We present a novel Active Magnetic Shield (AMS), designed and implemented for the n2EDM experiment at the Paul Scherrer Institute. The experiment will perform a high-sensitivity search for the electric dipole moment of the neutron. Magnetic-field stability and control is of key importance for n2EDM. A large, cubic, 5\,m side length, magnetically shielded room (MSR) provides a passive, quasi-static shielding-factor of about $10^5$ for its inner sensitive volume. The AMS consists of a system of eight complex, feedback-controlled compensation coils constructed on an irregular grid spanned on a volume of less than 1000\,m$^3$ around the MSR. The AMS is designed to provide a stable and uniform magnetic-field environment around the MSR, while being reasonably compact. The system can compensate static and variable magnetic fields up to \SI{\pm 50}{\micro \tesla} (homogeneous components) and \SI{\pm 5}{\micro\tesla / \meter} (first-order gradients), suppressing them to a few \SI{}{\micro \tesla} in the sub-Hertz frequency range. 
%While the presented system is of particular interest for high-sensitivity EDM searches, it could also be 
The presented design concept and implementation of the AMS fulfills the requirements of the n2EDM experiment and can be useful for other applications, where  magnetically silent  environments are important and spatial constraints inhibit simpler geometrical solutions.

\end{abstract}

\section{Introduction}
\label{sec:intro}
High-precision measurements in fundamental physics, using particles, nuclei, atomic, or molecular systems,  require exquisite temporal stability and spatial uniformity of many environmental parameters to control systematic effects and fully exploit their statistical sensitivity.  The control of the magnetic field is of particular importance in those experiments sensitive to the coupling of the magnetic field to the spin of the system through its magnetic moment. For example, experiments searching for permanent or variable electric dipole moments (EDMs), signals of dark matter fields, neutron-antineutron and mirror-neutron oscillations, Lorentz invariance violation, or new forces~\cite{pignol2019,Alarcon2022,Bitter1991,Safronova2018,Mirrors2021}. 
Most of them deploy dedicated coil systems generating uniform magnetic fields inside magnetically shielded volumes. Shielding of these volumes can be achieved by means of passive or active magnetic shielding (AMS), separately, or, in combination. Passive shields are built from high-permeability materials and rely on their magnetic properties. Active magnetic shields are based on feedback-controlled coils, where magnetic sensors detect changes of the magnetic field, and an algorithm calculates the proper response to adjust the coil currents and counteract the perturbation. 

Since the 1980s, numerous active shields have been built for different applications~\cite{SFC2014,Brake1991,Kelha1982,Voigt2013,Spemann2003,Brys2005,Kobayashi2012}, covering a wide range of research areas such as ion beams, electron microscopes, and bio-medical applications, as well as high-precision measurements of EDMs~\cite{Jungmann2013,Chupp2015PR,pignol2019,Alarcon2022}.
In particular, an active magnetic shield was successfully used for the first time by our collaboration in the nEDM experiment, which provides the current best measurement of the neutron EDM~\cite{nEDM-PhysRevLett}. The system consisted of six actively-controlled  rectangular coils with size of approximately 8~m $\times$~6~m, located in a Helmholtz-like positioning. The coils were built around a control volume of  2.5~m~$\times$~2.5~m~$\times$~3~m. The system was crucial to fully exploit the statistical sensitivity of the experiment~\cite{SFC2014}.

\begin{figure}[ht]
\centering
\includegraphics[width=0.45\textwidth]{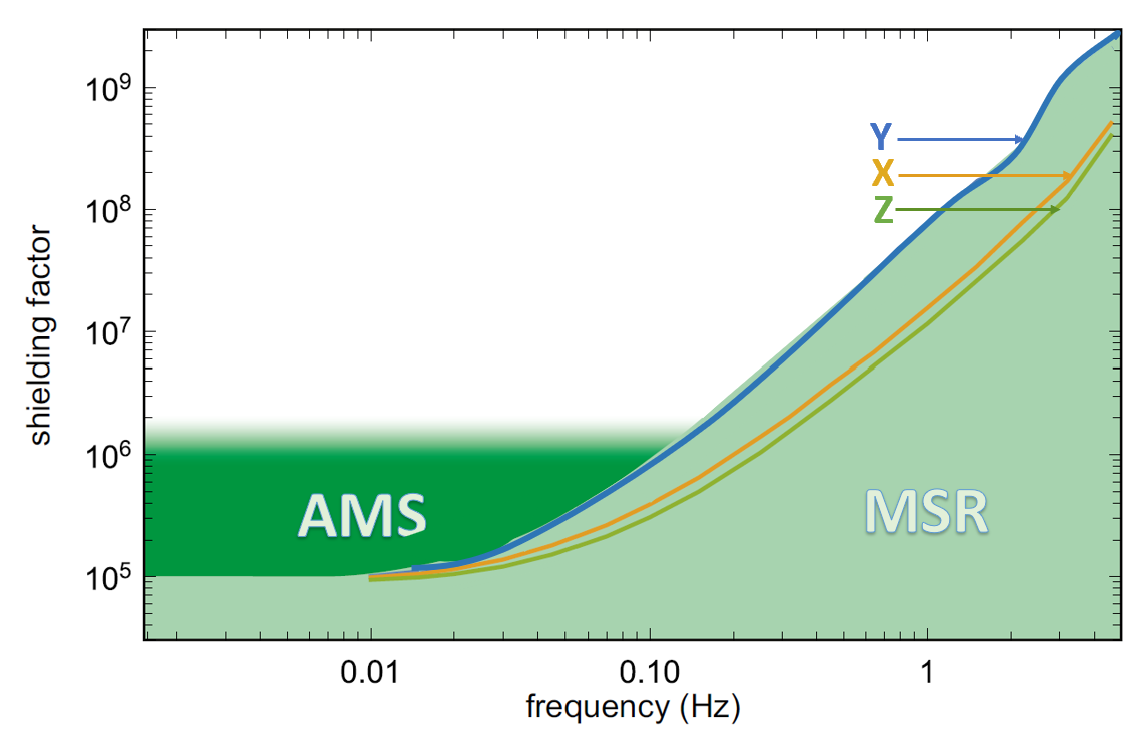}
\caption{\label{fig:AMS_MSR} 
Frequency-dependent shielding-factors of the MSR. The light green region and the colored curves for each spatial direction were obtained using external reference excitation coils to produce 2\,\SI{}{\micro \tesla} peak-to-peak sinusoidal fields at the central position of the MSR final location prior to the MSR assembly. The dark green region is the expected improved shielding-factor provided by the AMS system for large disturbances of order several 10\,\SI{}{\micro \tesla} in the low-frequency range. Adapted from~Ref.~\cite{MSR2022}.}
\end{figure}

In this paper, we report on the design-path, implementation, and initial performance characterization of a dedicated AMS for the n2EDM experiment~\cite{n2EDM,Rawlik2018PhD,Solange2021}, currently undergoing commissioning at the ultracold neutron (UCN) source~\cite{Lauss2022,Bison2022,Lauss2012} at the Paul Scherrer Institute (PSI). 
A tenfold improvement in statistical sensitivity of n2EDM over nEDM will be realized by many innovations, primarily by improved adaption to the UCN source and two enlarged verti\-cally-stacked UCN storage-chambers. The target systematic error budget yields stringent requirements for the magnetic-field stability and uniformity, and, thus, advanced shielding from magnetic-field disturbances. The n2EDM experiment uses a combination of passive and active shieldings around the sensitive volume. 
The passive shielding is provided by a Magnetically Shielded Room (MSR) ~\cite{MSR2022} with a base size of 5.2\,m x 5.2\,m and a height of 4.8\,m. It is composed of five mu-metal layers, one ULTRAVAC layer, and one intermittent RF-shielding layer with a shielding factor of $10^5$ at \SI{0.01}{\hertz} and rising with frequency to $10^8$ at \SI{1}{\hertz}, as shown in Fig.~\ref{fig:AMS_MSR}. 

The specifications for the n2EDM internal magnetic field are discussed in detail in~Ref.~\cite{n2EDM}. A crucial requirement is that the average magnitudes of magnetic fields in the two UCN storage chambers (vertically separated by 18\,cm) must not differ by more than 10\,pT, which corresponds to a vertical magnetic-field gradient smaller than 0.6\,pT/cm. This limits the systematic effects coming from the different resonance conditions in the two chambers.
%This condition guarantees that the spin ensembles in the two chambers share the same resonance conditions.

%In case external field changes would also not offset internal fields by more than 10\,pT,  
%the gradient condition could be fulfilled and, additionally, spin-flip frequencies would not have to be adapted to changing magnetic fields.

%
Due to the quasi-static shielding factor of the MSR of $10^5$, slow external field changes of order 1\,\SI{}{\micro \tesla} will lead to internal field changes of order 10\,pT. As the shell structure of the MSR and its high-quality, innermost layer tend to homogenize magnetic field changes, internal magnetic-field gradients resulting from external gradients are further suppressed~\cite{Bidinosti2014}. Thus, an external, inhomogeneous variation of a few \SI{}{\micro \tesla} around the MSR can be tolerated. However, larger field variations on the outside of the MSR could cause larger-than-allowed internal gradients. In addition, larger external field changes can change the magnetization of the outermost mu-metal layer of the MSR, which will, in turn, slowly propagate through the MSR layers, and result in undesirable drifts of the inner magnetic field. 

The task for the AMS in n2EDM is thus to provide a magnetic field around the MSR that is stable to within a few \SI{}{\micro \tesla}, even with sub-Hertz external variations, in order to meet the \SI{10}{\pico \tesla} conditions on the inside. For large, slow magnetic field variations, of order ten or several tens of \SI{}{\micro \tesla}, this also corresponds to
an improved overall shielding performance in the low-frequency regime,  see Fig.~\ref{fig:AMS_MSR}.

This paper describes the design and implementation of the AMS system for the n2EDM experiment~(see~\cite{Rawlik2018PhD,Solange2021})
and is organized as follows:

%{\bf in here: link to sections in the relevant sequence, perhaps need to reshuffle some content:}

(i) The magnetic fields over the complete volume occupied by the entire experimental apparatus were mapped before setting up the n2EDM experiment and are described in Sec.~\ref{sec:Mapping}. The disturbance of the field resulting from neighbouring magnetic instruments was evaluated. All relevant fields could be reproducibly measured and described to \SI{}{\micro \tesla} precision by superpositions of homogeneous (three directions) and first-order gradient (five independent components) magnetic-field contributions. This established the need for eight independent and ideally `orthogonal' coils for the field compensation system.

(ii) A method was developed to design optimal coils for specific magnetic fields when constraining the current-carrying wires to a predetermined, irregular grid on a surface around the volume of interest~\cite{Rawlik2018PhD,Rawlik2018}, described in Sec.~\ref{sec:AMS-concept}.

%Eight coils for the independent and orthogonal magnetic fields of the first two multipole orders (3 homogeneous and 5 first-order gradients) needed to be designed. 

%
(iii) A scaled-down prototype was developed and served as a proof-of-concept system, see Sec.~\ref{sec:Prototype-intro}. It allowed tests of various design options, including an irregular geometry, the powering of the coils, and the implementation of feedback sensors and appropriate algorithms, with and without mu-metal. 

(iv) A scheme to systematically simplify the individual, optimal, full-scale coils was developed to ease practical construction of the AMS without sacrificing the specified performances~\cite{Solange2021} (Sec.~\ref{sec:AMS-coil-design}). This included reducing windings in the eight coils and their efficient powering with eight current sources, each feeding three circuits. 

(v) The system was constructed with careful quality control during assembly of the system with more than 55\,km of cabling, as described in~ Sec.~\ref{sec:Technical-implementation}. 

(vi) Current sources were developed and implemented (Sec.~\ref{sec:Amplifiers}). An array of three-axis fluxgate sensors was implemented to monitor the magnetic field and inform the feedback algorithm (Sec.~\ref{sec:Fluxgates}). 

(vii) The commissioning of the full AMS system was successfully completed with various performance studies, as described in~Sec.~\ref{sec:AMS-performace-intro}.

\begin{comment}
The AMS design was based on an earlier-developed method~\cite{Rawlik2018} and driven by the sensitivity requirements in the n2EDM spectrometer taking into consideration spatial constraints  of the experimental setup.

In the following chapters we present the main concept of the AMS design, a small-scale proof-of-principle prototype and the implementation of the AMS system into the n2EDM experiment. 
\end{comment}

\section{Mapping of the experimental area}
\begin{figure}[ht]
\centering
\includegraphics[width=0.48\textwidth]{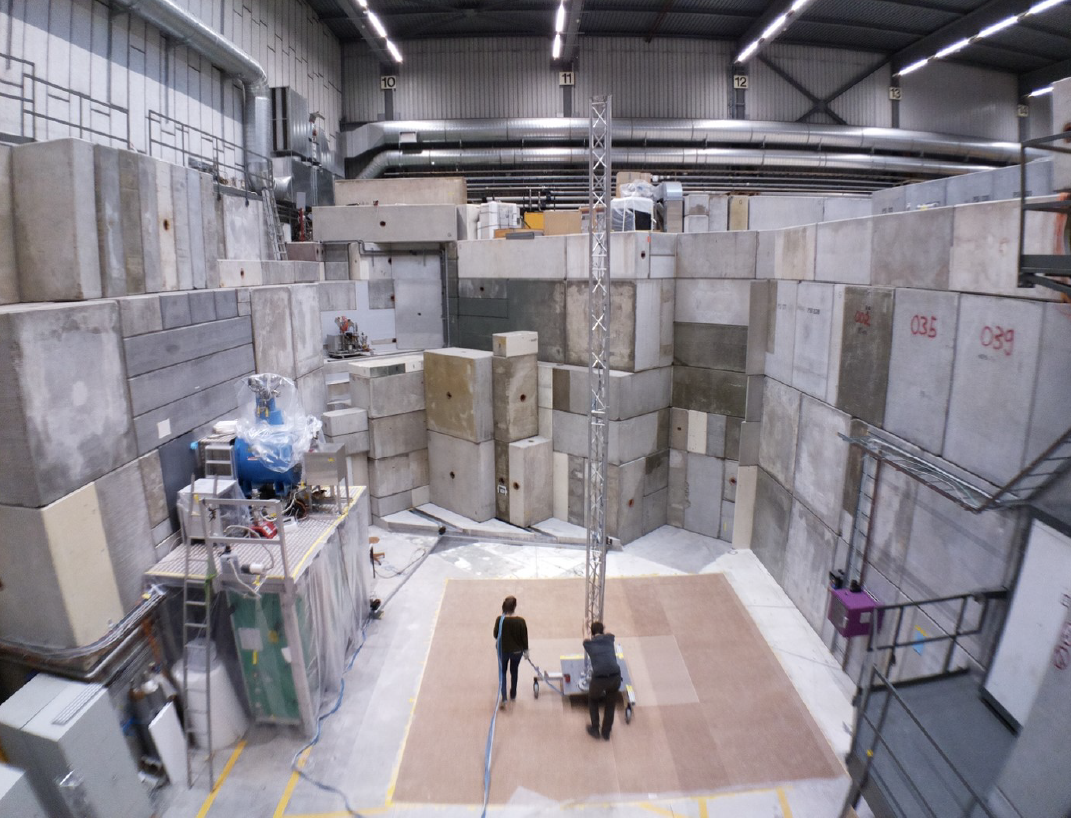}
\caption{\label{fig:MappingAS} Picture taken during a magnetic-field mapping in UCN area South at PSI~\cite{Solange2021}. The area was emptied before the n2EDM experiment was set up. Two people are moving the `mapping tower' around. About half height of the tower in the displayed position marks the center of n2EDM. The UCN source is behind the concrete shielding to the left, and a superconducting magnet (blue) to fully polarize UCN directed to n2EDM, is visible on the platform. See main text for further details.}
\end{figure}

%\label{sec:4.1}
\label{sec:Mapping}
The n2EDM experiment is located at PSI in UCN area South. Before setting up n2EDM, its predecessor nEDM was disassembled and the area cleared. Figure~\ref{fig:MappingAS} shows a view of the empty experimental area. The concrete blocks are part of the biological shield of the UCN source (to the left) and of the medical cyclotron COMET (forward direction and to the right). These blocks cannot be moved, and thus ultimately limit the space available for the n2EDM experiment. The MSR was decoupled from the rest of the hall on its own foundation, which is seen in the picture as brown floor, indicating approximately the size of the MSR base. Given the size of the MSR and the spatial constraints of the biological shields, the coils of the AMS system have to be as close as around 1 m to the MSR, and still providing the desired homogeneous field in the volume of interest. This immediately excludes AMS field generation with simple Helmholtz-like coil systems.

Before designing the AMS system,  the magnetic field of the experimental area was extensively mapped~\cite{Rawlik2018PhD,Solange2021} to determine the components of the magnetic field that have to be compensated. The measured field was then decomposed, by a least-squares fit, into zeroth order, homogeneous fields, first-order gradients, and higher-order contributions, obtaining field maps and interpolated continuous fields. 

Figure~\ref{fig:MappingAS} shows  the mobile `mapping tower' in action. The tower was constructed using  up to five identical, 2\,m-long aluminum triangle-truss segments from commercially available event stage equipment.  Each segment carried three 3-axis fluxgate sensors. The segments were vertically stacked and mounted on a heavy aluminum base plate on wheels. The position and orientation of this cart in the area was measured by three string potentiometers attached to a rigid coordinate system referenced to the area. The full area could thus be magnetically mapped within minutes, with spatial resolution limited by the reproducibility of the order of 0.1\,m.

Several strong superconducting magnets at 10\,m to 50\,m distances contribute with fields in the tens of \SI{}{\micro \tesla} range and field gradients of a few \SI{}{\micro\tesla}/m. Their influence is particularly severe as their fields can change during n2EDM measurements without prior notice. Typical time scale of such uncontrolled changes can vary from minutes to tens of minutes, possibly several times a day. So that the AMS can compensate those changes, for each of the known strong nearby magnets, the mapping was performed with it on and off. Thereby, the change of the magnetic field in the space of the n2EDM experiment was measured.

The superconducting magnet shown in Fig.~\ref{fig:MappingAS} is less problematic, as it is self-shielded with a known steep field gradient and controlled by the n2EDM experiment operation. During n2EDM operation it is always ramped up and  running very stably in a persistent mode. 

The reproducibility of maps taken under similar conditions was of the order of a few \SI{}{\micro \tesla}, where the limitation might be due to drifts of the fields themselves or uncertainties of the measurement and the analysis procedures. Importantly, it was found that the measured fields could be sufficiently described, with \SI{1}{\micro\tesla}-accuracy, with only homogeneous and first-order gradient fields.

We concluded that the AMS needed only coils to compensate homogeneous fields in the three independent spatial directions, and five independent first-order gradients. Therefore, a system could be designed using only eight independent coils. 

Concerning field strengths, it was found that a range of \SI{\pm 50}{\micro\tesla} for the three homogeneous components of the field, and up to \SI{\pm 5}{\micro\tesla / \meter} for the five first-order gradients would be sufficient to meet our requirements. These values already include a safety margin of 20\%.

%
%The magnetic field in the volume of interest was mapped in the empty experimental area. 
%Various nearby strong magnets from other research installations were ramped on and off during the mapping campaign. 
%
%In order to compensate their fields, it was found necessary to achieve a range of \SI{\pm 50}{\micro\tesla} in all three homogeneous components of the field, and all five first-order gradients of up to \SI{\pm 5}{\micro\tesla / \meter}.
%
%This takes a 20\% safety margin into account.
%
%This defined the target fields of the AMS system design. Second and higher order gradients were sufficiently small that they did not need specific coils. 
%
%It was also measured that the fields produced by these magnets are reproducible to a level of few~\si{\micro\tesla}. 

\section{The concept of the AMS design}
\label{sec:AMS-concept}
% The goal of the AMS system presented in this paper is to provide a stable and uniform magnetic environment around the MSR of the n2EDM experiment, which defines the main features of its design. In this chapter we present the concept of the AMS design followed by its prototyping at ETH Z{\"u}rich.

%\subsection{Goals of the AMS system}
%%\label{sec:2.1}
%\label{sec:AMS-goals}
%\input{AMS-goals}

\subsection{Working principle}
\begin{figure}
\centering
\includegraphics[width=0.5\textwidth]{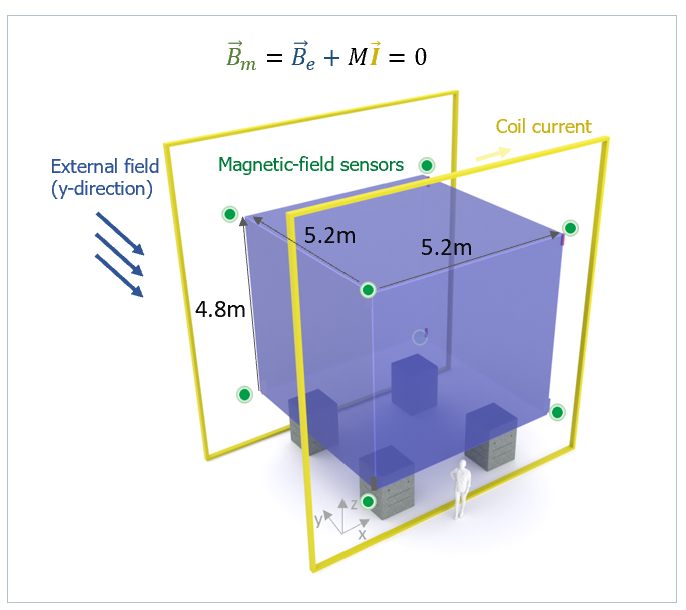}
\caption{\label{fig:AMSn2EDM_principle} The volume of interest for the target fields of the n2EDM AMS system are given by the outside wall of the MSR (depicted in violet). An external magnetic field $\boldsymbol{B}_{\mathbf{e}}$ is detected by magnetic-field sensors (green). As an example, the yellow coils could aim to compensate the external field. In practice, the AMS coils are more complicated due to spatial limitations and their close proximity to the MSR.}
\end{figure}

%\label{sec:2.2}
\label{sec:Working-principle}
%The working principle of the active magnetic shielding is based on the compensation of the environmental magnetic field and the magnetic-field distortions by actively-controlled coil currents which counteract the magnetic-field changes in the controlled volume. These changes are detected by dedicated magnetic-field sensors. Their readings are used in a feedback loop to drive the coil currents, see Fig.~\ref{fig:AMSn2EDM_principle} for a schematic. 
%
In a volume with no magnetised parts, any magnetic field can be generated by the correct current distribution on the surface of this volume. The currents on the surface can thus be chosen to exactly counteract the effect of any external field, stabilising the field inside. In a practical realisation, there is a finite number of coils on the surface, and the field is measured only in a finite number of points.  Figure~\ref{fig:AMSn2EDM_principle} depicts a simple realisation with a single coil (shown in yellow) and eight 3-axis sensors (shown in green).
 Here, an external magnetic field $\boldsymbol{B}_{\mathbf{e}}$ influences the target volume, in which the field is to be stabilised (depicted in blue, occupied by the MSR of n2EDM). We aim, however, for optimal (in the least-squares sense) stabilization at the eight green points where three-axis sensors are placed. Their readings $\boldsymbol{B}_{\mathbf{m}}$ are used to calculate currents $\boldsymbol{I}$ feeding a coil system  to counteract external field changes. Obviously, in the real application, the coils are much more complicated than shown. In principle one can aim at any target field at the surface of the sensitive volume. In our application aiming at zero field is most reasonable.

In the absence of the MSR and other magnetization, one obtains a linear dependence between $\boldsymbol{B}_{\mathbf{m}}$ and the coil currents. 
In fact, we initially assume, and later prove experimentally in section \ref{sec:AMS-shielding}, that a linear dependence also holds for a demagnetized MSR exposed only to small magnetic fields. One can write:
\begin{equation}
\boldsymbol{B}_{\mathbf{m}} =\boldsymbol{B}_{\mathbf{e}}+\boldsymbol{M}\,\boldsymbol{I}.
\label{Eq:basics}
\end{equation}
The matrix  $\boldsymbol{M}$ contains the proportionality factors, which relate the current in the AMS coils to the magnetic fields measured at the sensor positions. For a built system, the entries of $\boldsymbol{M}$ can be measured using the installed coils and sensors (see Sec.~\ref{sec:Prototype-Performance}). During design of the coil itself, they can be calculated, without the MSR  using Biot-Savart's law, and with the MSR, using a sufficiently realistic finite-element simulation of the full system.  Equation~\ref{Eq:basics} can be written components-wise in the following way:

\begin{equation}
\begin{pmatrix}
B_{{\mathrm{m}},\text{1x}}\\
B_{{\mathrm{m}},1\text{y}} \\
B_{{\mathrm{m}},1\text{z}} \\
\vdots \\
B_{{\mathrm{m}},n\text{z}}
\end{pmatrix} =
\begin{pmatrix}
B_{\mathrm{e},1\text{x}}\\
B_{\mathrm{e},1\text{y}}\\
B_{\mathrm{e},1\text{z}}\\
\vdots\\
B_{\mathrm{e},n\text{z}}
\end{pmatrix} 
+
\begin{pmatrix}
     M_ {11\text{x}} & M_ {21\text{x}} & \dots & M_ {k1\text{x}}\\
     M_ {11\text{y}} & M_ {21\text{y}} & \dots & M_ {k1\text{y}}\\
     M_ {11\text{z}} & M_ {21\text{z}} & \dots & M_ {k1\text{z}}\\
     \vdots & \vdots & \ddots & \vdots \\
     M_ {1n\text{z}} & M_ {2n\text{z}} & \dots & M_ {kn\text{z}}\\
   \end{pmatrix}
\begin{pmatrix}
I_ {1}\\
I_ {2} \\
\vdots \\
I_ {k}
\end{pmatrix}.
\label{Eq:basics-components}
\end{equation}
Matrix $\boldsymbol{M}$ has dimensions $3\,n~\times~k$, where $k$ is the number of coils (for the AMS system $k=8$, see Sec.~\ref{sec:Method}), and $n$ is the number of magnetic-field sensors.

Next, one implements an iterative process using the measured changes in $\boldsymbol{B}_{\mathbf{m}}$ (see Sec.~\ref{sec:ControlSystem}) to calculate appropriate changes for $\boldsymbol{I}$ to zero $\boldsymbol{B}_{\mathbf{m}}$ again.
In the applied feedback algorithm the pseudo-inverse of $\boldsymbol{M}$ is used. 
In order to calculate the pseudo-inverse we start with a Singular Value Decomposition:
%unused? %(SVD):
%
\begin{equation}
\boldsymbol{M}=\boldsymbol{U}\boldsymbol{S}\boldsymbol{V}^{\text{T}},
\label{Eq:SVD}
\end{equation}
where $\boldsymbol{U}$ and $\boldsymbol{V}$ are unitary matrices and $\boldsymbol{S}$ is diagonal. The latter is called the spectrum and describes the effect of combinations of coils on the magnetic sensors. The pseudo-inverse $\boldsymbol{M}^{-1}$ can be calculated as:
\begin{equation}
\boldsymbol{M}^{-1}=\boldsymbol{V}\boldsymbol{S}^{-1}\boldsymbol{U}^{\text{T}}.
\label{Eq:Inverse}
\end{equation}
The ratio of extreme values of the spectrum, $s_{min}$ and $s_{max}$, defines a \textit{condition number C} of the matrix $\boldsymbol{M}$:
\begin{equation}
C=\frac{s_{max}}{s_{min}}.
\label{Eq:condition}
\end{equation}
The condition number is an important characteristic of the system design of the feedback-matrix quality. It represents the sensitivity of the sensors to current changes. A low condition number means that there are particular combinations of currents I that have small influence on the readings of the sensors measuring $\boldsymbol{B}_{\mathbf{m}}$. When inverted, this leads to small changes in the sensors to cause large changes in the currents, rendering the system to be unstable.
The condition number is later used in the optimization of the sensor positions in the AMS system (Sec.~\ref{sec:Prototype-intro}).

\subsection{Design challenges}
%\label{sec:2.3}
\label{sec:Challenges}
In a volume with no magnetized parts, any magnetic field can be generated by the correct current distribution on the surface of this volume. In particular, the currents on the surface can be chosen to exactly counteract the effect of any external field, making the inner magnetic field zero. The MSR can be demagnetized and reside in the zero field inside the volume with exactly the same currents as needed for the empty volume without the MSR. 

In the real experiment, currents cannot be arbitrarily distributed on a surface. They must follow predefined, discrete paths and the fields can only be adjusted by varying current values. The spatial discretization is a grid to which the current carrying wires are fixed. 
The discretized current distribution can approximate the target field well, if the discretization in small compared to the distance between the target volume and the surface. 

Several constraints for the AMS system were already mentioned. The coil system must be large compared to the size of the MSR, however, the walls of the experimental area ultimately limit the size of the surface to which the AMS could be mounted. In addition, the experimental area must be accessible. It should be possible for persons with reasonably sized equipment to enter the experiment without breaking the currents in the AMS. It should also be possible to open the AMS from the top to insert large equipment with the crane. Various other installations penetrate surfaces around the MSR such that the grid for the AMS must be adapted to the needs of other subsystems. 

The shielding blocks seen in Fig.~\ref{fig:MappingAS} as well as the regular floor of the experimental hall are made of steel-reinforced concrete with some magnetic response. The latter was investigated and fortunately found to be rather weak and finally negligible if a minimal distance to the walls is maintained.
%

\begin{comment}

The design goal of the AMS system was to compensate environmental fields, so that the field around the MSR outside wall is stable and close to zero~\cite{n2EDM}.
%
Fig.~\ref{fig:AMSn2EDM_principle} schematically displays the volume of interest for the n2EDM AMS system as given by the MSR. 
%
As an example, depicted yellow coils could aim to compensate the external field. The size of the coils is limited by the surrounding experimental infrastructure.
%
The minimal distance between the compensating coil and the MSR walls does not exceed 1.5m. 
%
Due to these limitations,  we used the method \cite{Rawlik2018} to develop more complicated AMS coils, to produce homogeneous field as required in the volume of interest.

\begin{figure}
\centering
\includegraphics[width=0.5\textwidth]{Images/Y-coil-new.png}
\caption{\label{fig:AMSn2EDM_principle} Scheme to display the volume of interest for the n2EDM AMS system as given by the outside wall of the MSR (depicted in violet). An external magnetic field ${B_{e}}$ is detected by magnetic-field sensors (green). As an example, the yellow coils could aim to compensate the external field. In practice, the AMS coils are more complicated due to spatial limitations and their close proximity to the MSR.}
\end{figure}

  This method allowed to design coils with wire paths along a predetermined grid that is mounted on the inside of the thermal shell around the MSR. In the next section we briefly introduce the basic principles of the method.

\end{comment}

\subsection{Method of simple coil design and its application for the AMS system}
%\label{sec:2.4}
\label{sec:Method}
The aforementioned requirements and constraints of the system called for the development of a flexible method to design coils that could be practically built.

\begin{figure*}
\centering
    \includegraphics[width=0.7\textwidth]{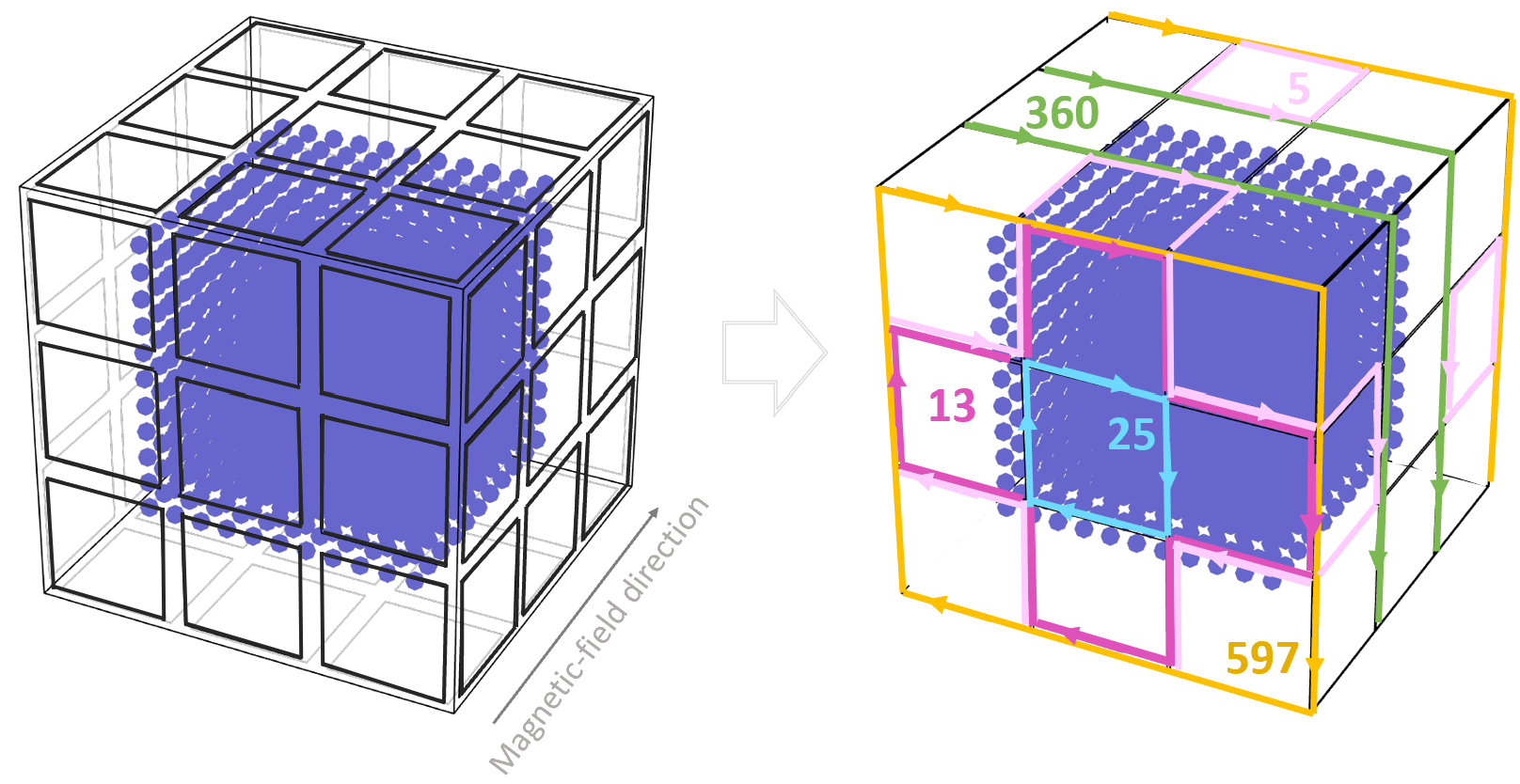}
  \caption{Illustrations of the method of simple coil design ~\cite{Rawlik2018}. Left: Initial definition of the grid (bold tiles) around points of interest (blue dots). Right: a set of simple loops (different colors) obtained in the course of simple coil design, representing the current paths needed to create the desired field (here: homogeneous in the indicated direction). Numbers and arrows indicate values of the currents and their directions in the corresponding loops. More details can be found in~\cite{Rawlik2018PhD}.}
\label{fig:coil-method}
\end{figure*}

The employed method of coil design ~\cite{Rawlik2018,Rawlik2018PhD} is based on three key inputs: 
(i) target fields, which have to be compensated by the coils; 
(ii) a fixed grid where coil wires can be placed; and 
(iii) points of interest (POIs) of target fields, covering the fiducial volume of interest densely enough (Fig.~\ref{fig:coil-method}, left). 
The grid can be subdivided into many small coils called tiles. A tile is the smallest building block of the grid. The smaller this elementary building block is, the more homogeneous the field can be. For practical reasons, we have chosen to make most tiles rectangular; however, the method could deal with any shape.
It is also worth reiterating that the method does not require the grid to be regular. In fact,  the AMS system of the n2EDM experiment (Sec.~\ref{sec:AMS-n2EDM-into}) is implemented on an irregular grid.

Once target fields, grid, and POIs are defined, the magnetic field at the POIs,
$\boldsymbol{B}_{\mathbf{POI}}$, created by the currents $\boldsymbol{I}$ in the tiles, can be described similarly to Eq.~\ref{Eq:basics} by
\begin{equation}
\boldsymbol{B}_{\mathbf{POI}} = \boldsymbol{M}_{\mathbf{D}}\,\boldsymbol{I}.
\label{Eq:method}
\end{equation}
In this design phase, each element of proportionality matrix $\boldsymbol{M}_{\mathbf{D}}$ can now be calculated numerically using Biot-Savart's law.
Using a least-squares method we find the current needed in each tile to approximate the target field. 
For the AMS system, there were 308 tiles in total (see~Sec.~\ref{sec:AMS-coil-design}).
The calculation is simplified by cancelling counteracting currents on the grid structure. 
The algorithm described in \cite{Rawlik2018PhD} decomposes this grid of currents into simple loops, which are closed current paths that can be wound on the grid. 
The result of this step is a set of such loops, that each need to be powered with a specific design current to generate the target field. 
Examples of such simple loops are depicted in different colors in the right panel of Fig.~\ref{fig:coil-method}. 

In order to change the magnitude of the field generated by a system of loops, all loop currents in the system need to change proportionally to the design current. 
Thus, the set of loops for a specific target field can be connected in series, creating one \textit{coil}. 
The number of windings for each loop can be adjusted, such that the coil could be powered by one current source.
%power supply. 
%
However, for the AMS system it was decided to split each coil into three electrical circuits: with large, medium and small elementary currents (which will still be changed with the same proportionality and are integer multiples of the smallest current). %
For example, choosing elementary currents as [15A, 5A, 1A], a loop with a current of $\SI{73}{\ampere}$ would be wound as follows: 
\begin{equation}
\SI{73}{\ampere}=4\times\SI{15}{\ampere}+2\times\SI{5}{\ampere} + 3\times\SI{1}{\ampere}.
 \label{eqn:winding}
\end{equation}

The choice of the smallest elementary current leads to some imperfection, here on the order of $0.5/73 \approx 0.7\%$ or 
$\SI{0.4}{\micro \tesla}$ for $\SI{50}{\micro \tesla}$, within the requirements of the system.
This approach allowed minimization of winding efforts and self-inductance, while keeping the number of current source channels reasonable. In our case, we constructed eight independent coils, each with three circuits for the elementary currents. They are operated by eight current sources, each with three channels for the different currents.

The described method of coil design for an AMS offers advantages over approaches using simple geometric coils, namely in two areas. 1) The size of the coil system can be decreased relative
to the size of the sensitive volume. A loss of performance, e.g., in the
homogeneity of a given volume, can always be counteracted by choosing a
denser grid.
2) The method allows construction of a coil for any field and the grid geometry that can be chosen almost arbitrarily. In particular, we have chosen to construct coils that produce orthogonal fields. The magnetic field is described as a superposition of orthogonal, cartesian harmonic polynomials:

\begin{equation}
	\mathbf{B}(\mathbf{r}) = \sum_{n=1}^{n_{\text{max}}} H_n \, \mathbf{P}_n(\mathbf{r}).
 \label{eqn:principle}
\end{equation}

\noindent Here $H_n$ are expansion coefficients and  $\mathbf{P}_n(\mathbf{r})=(P_{nx},P_{ny},P_{nz})$ are polynomials as defined in Table~\,\ref{tab:AMS-Cartesian-harmonic-polynomials} for the three homogeneous  %(n=1 (x) -3 (z)$) 
and five first-order gradient %($n=4 (G_1) - 8 (G_5)$) 
fields. 
%Each basis state already satisfies Maxwell’s equations. 
%The first three terms are homogeneous fields, the next five are the five independent linear gradients. 
Terms of higher $n$ correspond to
higher-order gradients. The advantages of this decomposition are that the polynomials are orthogonal and each basis state satisfies Maxwell's equations. This method was initially considered by G.~Wyszynski~\cite{Wyszynski2017}. We used this approach to define eight AMS coils: three coils to compensate homogeneous fields, and five to compensate linear magnetic field gradients.

\begin{table}
	\renewcommand{\arraystretch}{2}
	\centering
	\begin{tabular}{|c|c|c|c|c|}
		\hline	
Coil& $n$	& ${P_{n}}_{x}$	& ${P_n}_y$	& ${P_n}_z$ \\
\hline	
x& 1		& 1	& 0	& 0 \\
y& 2		& 0	& 1	& 0 \\
z& 3	    & 0	& 0	& 1 	\\
\hline	
G$_1$& 4		& x	& 0	& -z \\
G$_2$& 5		& y	& x	& 0 \\
G$_3$& 6		& 0	& y	& -z \\
G$_4$& 7		& z	& 0	& x \\
G$_5$& 8		& 0	& z	& y \\
\hline	
\end{tabular}
\caption[]{List of the Cartesian harmonic polynomials and associated names of the individual coils.}
\label{tab:AMS-Cartesian-harmonic-polynomials}
\end{table}

%In the next section we describe a first proof-of-principle prototyping of such a system.

\section{The AMS prototype}
%\label{sec:3}
\label{sec:Prototype-intro}
Before applying the simple-coil method to design and construct the AMS system for n2EDM, we studied a smaller-scale prototype~\cite{Rawlik2018PhD,Solange2021} at ETH Zurich~(Fig.~\ref{fig:prototype}). 
%It has been done to a large extent within the PhD theses of M.~Rawlik~\cite{Rawlik2018PhD} and S.~Emmenegger~\cite{Solange2021}.

\subsection{The prototype design and construction}
%\label{sec:3.1}
\label{sec:Prototype-main}
The prototype consisted of the eight types of coils as intended for the AMS system.
%(as described in Sec.~\ref{sec:AMS-n2EDM-into}). 
Similar in specifications, the system was designed to compensate target fields of \SI{\pm 50}{\micro \tesla} for homogeneous fields, and \SI{\pm 20}{\micro\tesla / \meter} for the first-order gradients. It also aimed at a homogeneity of a few~\SI{}{\micro \tesla} in the volume of interest.

\begin{figure}[t]
\centering
\includegraphics[width=0.4\textwidth]{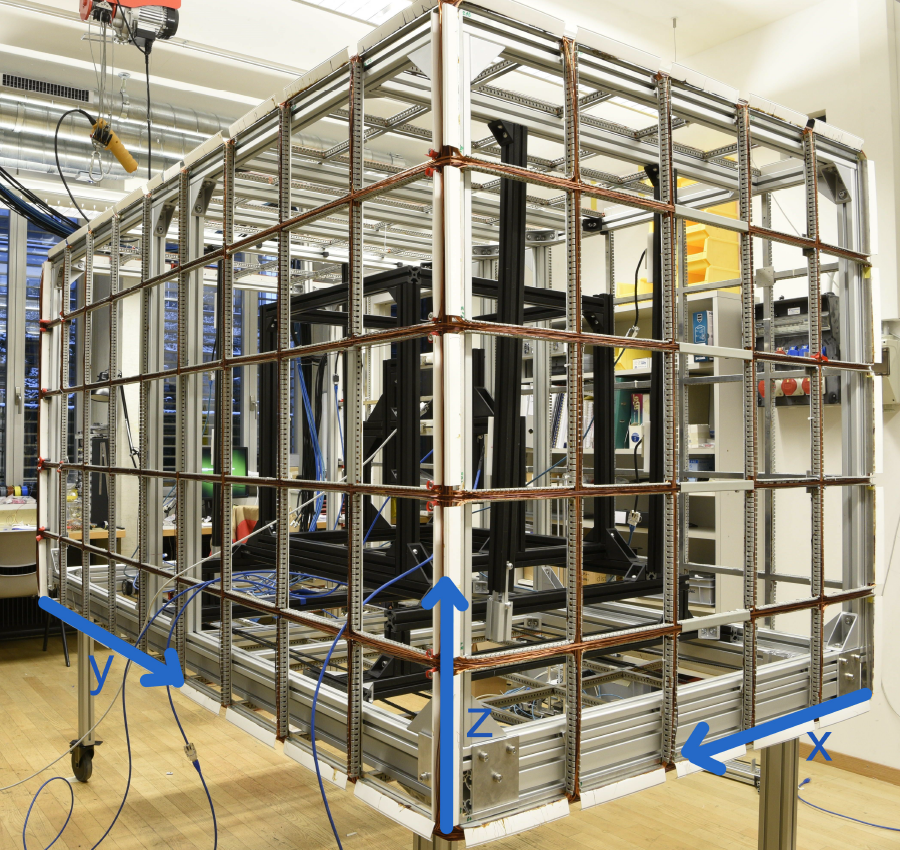}
\caption{\label{fig:prototype} Photo of the AMS prototype ~\cite{Rawlik2018PhD,Solange2021} mounted in the ETH laboratory. The smaller side of the frame, facing the window as seen on the photo, was kept open without windings - to allow easy access to the inside of the system.}
\end{figure}

\begin{figure}[t]
\centering
\includegraphics[width=0.35\textwidth]{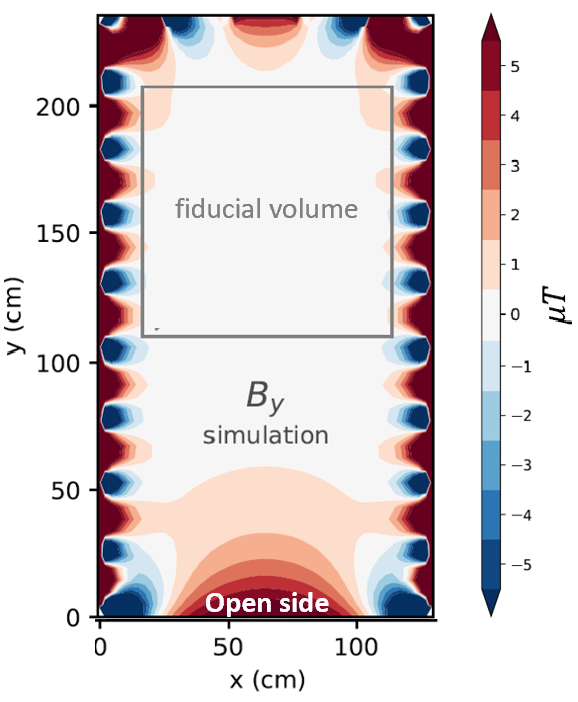}
\caption{\label{fig:prototype-simulation} 
Simulation of the y-component of the magnetic field produced by the y-coil of the AMS prototype in the x-y midplane. Shown are deviations of the magnetic field from the target value of \SI{50}{\micro \tesla}. The grey contour depicts the volume of a removable cubic mu-metal shield, which was not considered in the simulation. Figure is adapted from~\cite{Rawlik2018PhD}.
}
\end{figure}

The prototype was built on an aluminum-profile frame of 1.3\,m~$\times$~2.3\,m~$\times$~1.3\,m in $x$-, $y$-, and $z$-directions, respectively, providing a grid of squares as shown in Fig.~\ref{fig:prototype}.
The sensitive, fiducial volume for the target fields was chosen to be a  cube of 98\,cm~$\times$~98\,cm~$\times$~98\,cm, placed asymmetrically in $y$-direction and centered in $x$ and $z$, as shown in Fig.~\ref{fig:prototype-simulation} (grey contour), in which a cubic mu-metal shield could be placed. We kept the $x-z$ side of the frame at $y=0$ completely open, opposite to the front-side seen in Fig.~\ref{fig:prototype}. This enabled easy access to the inside, e.g., to install a magnetic-field mapping device and the mu-metal, and in addition demonstrated the feasibility of designing and building coils with a more complex, irregular geometry. 

The mu-metal cube in the prototype served as the emulation of the MSR of n2EDM concerning the fields on its outside. Its purpose was not to be an efficient magnetic shield but rather to provide a mu-metal surface to affect the fields between the mu-metal and the coil cage. It can be demagnetized using a set of demagnetization coils wound through the cube.

The design method restricted the wires of the coil system to take paths on the grid, similar to the ones shown in the right panel of Fig.~\ref{fig:coil-method}.
As described in Sec.~\ref{sec:Working-principle}, each of the coils used three circuits with different values of maximal currents (here: \SI{5}{\A}, \SI{1}{\A}, and \SI{0.2}{\A}).
In total, eight current sources, each feeding three circuits, were used to provide all eight coils with their currents.

As an example, Fig.~\ref{fig:prototype-simulation} shows a simulation of the $y$-com\-po\-nent of the magnetic field produced by the homogeneous $y$-field coil of the prototype in the $x-y$ midplane. The map depicts deviations of the magnetic field from the target value of \SI{50}{\micro \tesla}.  The designed and predicted homogeneity of the field did not exceed a few \SI{}{\micro \tesla} in the sensitive volume. Similar results were obtained for all the coils.
%
%In the next sections we present experimental validation of these simulations as well as measurements of the shielding performance of the AMS prototype.

\subsection{Performance of the prototype}
%\label{sec:3.2}
\label{sec:Prototype-Performance}

\begin{figure*}[t]
	\centering
	\includegraphics[width=0.85\textwidth]{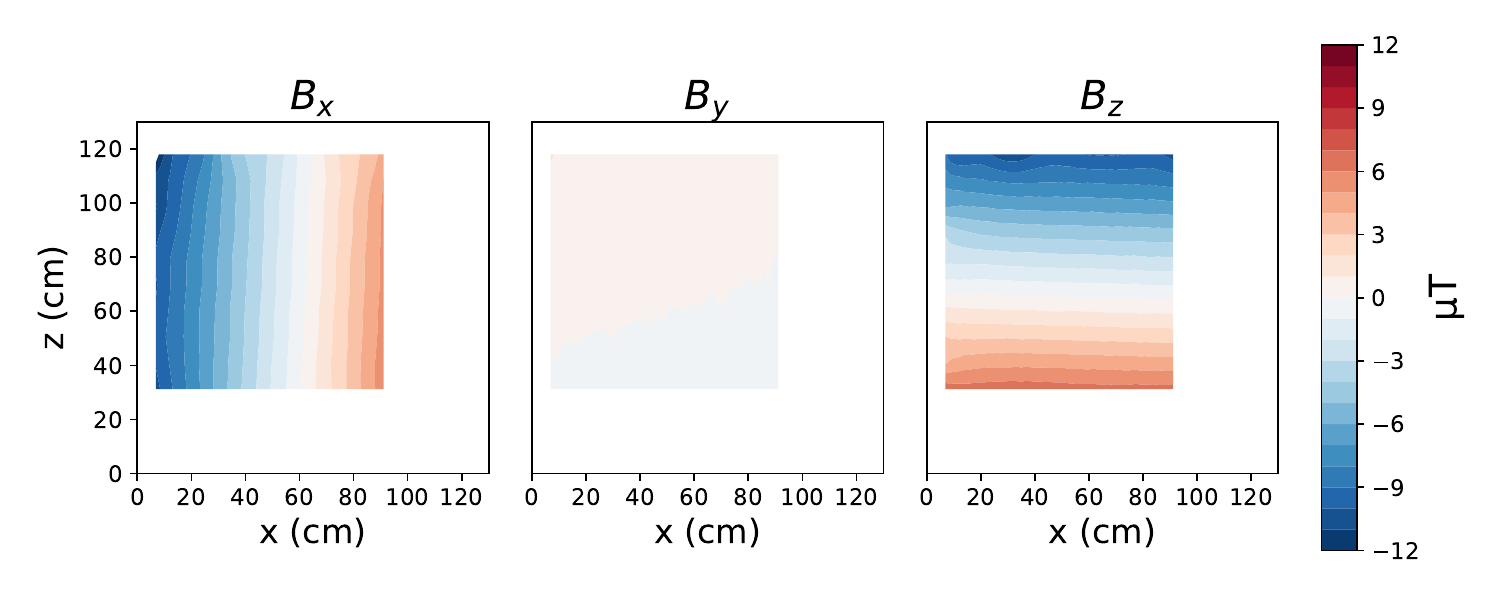}
	\caption{\label{fig:prototype-residuals} The map of the magnetic field produced by the first-gradient coil of the AMS prototype (G$_1$, as defined in~Table~\,\ref{tab:AMS-Cartesian-harmonic-polynomials}), measured at the central $x-z$ plane at $y = 115$\,cm. Adapted from~\cite{Rawlik2018PhD}.}
\end{figure*}

\paragraph{Validation of the AMS fields.}
We built a mapper robot carrying a movable three-axis fluxgate sensor to automatically measure the magnetic field in a large part of the volume inside the coil cage. 
In a first characterization, the static performance of the prototype was assessed by comparing the predicted and the measured fields for each coil. As an example, Fig.~\ref{fig:prototype-residuals} shows measurement results for the magnetic field produced by the first-gradient coil G$_1$ (as defined in~Table~\,\ref{tab:AMS-Cartesian-harmonic-polynomials}) at the central $x-z$ plane, at $y = 115$\,cm. As expected, the coil produces mainly $B_x$ and $B_z$ components of the field. The deviation of the measured fields from the target values do not exceed a few \SI{}{\micro \tesla}, which was the design goal and have been confirmed for all coils.

%\textbf
\paragraph{Dynamic field stabilisation.}
As a next step, we implemented a dynamic field stabilisation to actively suppress variable magnetic-field perturbations. 
This mode is based on continuous monitoring of the magnetic-field changes by fluxgate sensors with \SI{\pm 200}{\micro \tesla} range,  \SI{1}{\kilo \hertz} bandwidth, and \SI{\pm 0.5}{\micro \tesla} accuracy. The sensors were mounted around the volume of interest and a dedicated DAQ system, based on Beckhoff EtherCAT modules~\cite{Beckhoff}, was used to read their outputs, and to control the coil currents~\cite{Rawlik2018PhD}.

The dynamic mode of operation relies on the quality of the feedback matrix $\mathbf{M}$ (Eq.~\ref{Eq:basics}), which itself strongly depends on the number and positioning of the fluxgates, requiring optimization. 
The optimization without mu-metal is straightforward. It is somewhat more challenging with mu-metal due to its strong position-dependent impact on the magnetic field in its vicinity. 
The magnetic fields of the setup with the mu-metal cube were simulated with COMSOL~\cite{COMSOL} and validated by measurements.
With these simulations, the condition number (see Sec.~\ref{sec:Working-principle}) of the feedback matrix $\mathbf{M}$  could be minimized by selecting proper positions for the feedback sensors.

It was found that sufficiently stable performance can be reached with eight sensors placed close to the corners of the mu-metal. 
This fits well the intuitive understanding of the effects of mu-metal. Close to the surface of the mu-metal, field components parallel to the surface will be small while the orthogonal component remains. 
As the fluxgate magnetometers used for feedback are three-axis devices, this would mean that a fluxgate aligned near a large flat surface can measure the orthogonal field well with one axis, while two axes provide relatively little useful information. 
Thus, positioning fluxgates closer to mu-metal edges and  corners turns out to be more informative.

\begin{figure}[t]
\centering
\includegraphics[width=0.45\textwidth]{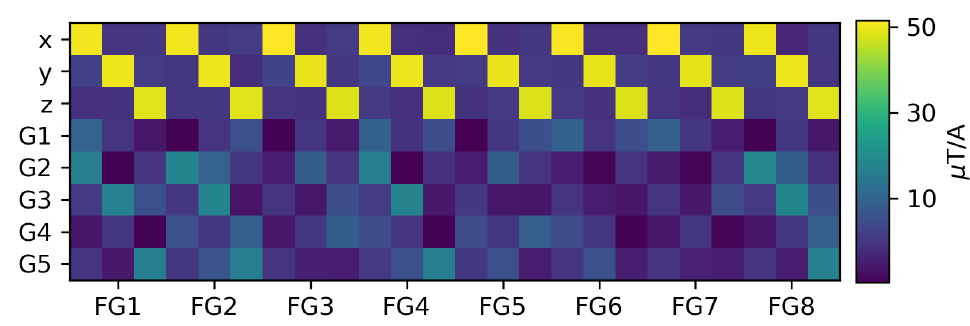}
\caption{\label{fig:matrix-prototype} Visualization of the feedback matrix $\mathbf{M}$ (see Eq.~\ref{Eq:basics} and~\cite{Rawlik2018PhD}). The rows designate the coils of the AMS
prototype and the columns correspond to the absolute readings in $x$, $y$ and $z$ directions of the eight fluxgates (FG).}
\end{figure}

With the fluxgates mounted at their optimal positions,  the feedback matrix was determined by measuring the mag\-net\-ic-field components while scanning each coil current separately over the whole available range. 
%%%%% verify: %%%%
For fields up to \SI{\pm 50}{\micro \tesla}, relevant here, the response was found to be linear.
%%%%%%%%%%%%%%%%%%
The slopes, obtained by linear regressions for each spatial field component versus current,
correspond to the elements of the matrix $\mathbf{M}$. They are displayed in Fig.~\ref{fig:matrix-prototype} for the eight 3-axis sensors and eight coils.

%\textbf{Shielding performance of the AMS prototype.}
\paragraph{Shielding performance of the AMS prototype.}
The obtained feedback matrix was  used to operate the AMS in dynamic-stabilization mode. To test this regime, magnetic-field perturbations can be generated using a coil placed at some location around the setup.
 For this measurement, the square excitation coil with sides around 1\,m was used. The coil was oriented perpendicular to the $x$ axis and placed at about $x=3$\,m with its center on the $y, z$ coordinate of the center of the sensitive volume.  The current source for the coil was modulated with a waveform generator to produce sinusoidal fields, with an uncompensated, maximal amplitude of about \SI{8}{\micro\tesla} at the central sensor position. The readout bandwidth was \SI{200}{\hertz} and the update rate of the feedback system was around \SI{30}{\hertz}.

%%%%%%%%%%%%%%%%%%%%%%%%%%
%%%%%%%%%%%%%%%%%%%%%%%%%%

The  AMS shielding performance was characterized by  a shielding factor $S$, defined as the ratio:

\begin{equation}
S = \frac {B_{\text{center}}^{\text{on}}}  {B_{\text{center}}^{\text{off}}},
 \label{eqn:shielding}
\end{equation}

\noindent
\\  where $B_{\text{center}}^{\text{on/off}}$ denotes the magnetic-field value at the center of the sensitive volume (corresponding also to the center of the mu-metal cube) with the active compensation on or off, respectively. The mu-metal cube was not used for the particular measurement explained here. In addition to the feedback sensors, one additional sensor to measure $B_{\text{center}}$ was mounted.

\begin{figure}[t]
\centering
\includegraphics[width=0.45\textwidth]{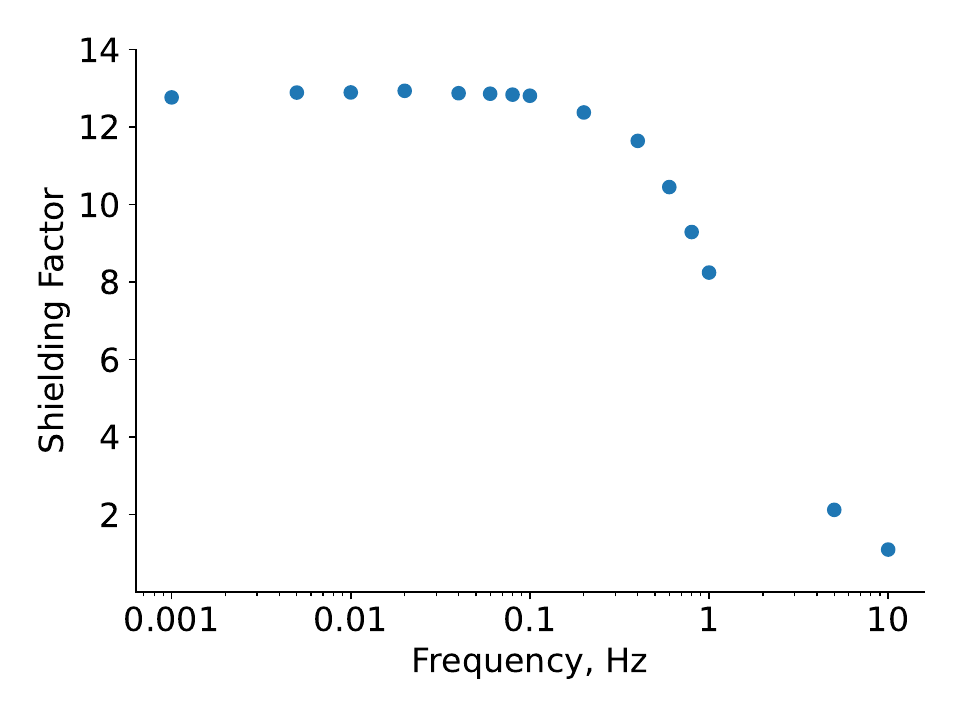}
\caption{\label{fig:prototype-shielding} Frequency-dependent shielding factor of the AMS prototype along its $x$ direction measured with the perturbation from a square-shaped coil with 1\,m side at a distance of about 3\,m, producing external sinusoidal fields of different frequencies. Adapted from~\cite{Solange2021}. See text for details. }
\end{figure}

Figure~\ref{fig:prototype-shielding} shows the obtained result for $S$ and its  frequency dependence up to \SI{1}{\hertz}. The measured shielding factor stays stable around 12 in the low-frequency range. The decrease of the shielding factor above \SI{100}{\milli \hertz} is from the limited response of the system caused by the inductance of the coils on the aluminum frame of the cage.
The frequency dependence of the shielding factor stays the same for similar measurements with the coil positioned at other locations. However, the magnitude of $S$ strongly depends on the distance and orientation of the excitation coil. This can be understood qualitatively, as the magnitude of the higher order gradients of the fields within the sensitive volume depends on distance and orientation of the coil, and cannot in principle be compensated by a first-order AMS. Also, this was quantitatively confirmed using a COMSOL simulation of the system.

In summary, we successfully demonstrated an implementation of the method of simple coil design with the prototype AMS system, achieving the expected static and dynamic performance. The prototype design with an open-side demonstrated that this approach is capable of handling irregular grids, which is important to account for doors and other openings at the n2EDM experiment.  Based on this feasibility demonstration, confidence was gained for the design and construction of the much larger AMS for n2EDM.

\section{The AMS system for n2EDM}
%\label{sec:4}
\label{sec:AMS-n2EDM-into}
%%%%%%%%%%%%%%%
%%%%%%%%%%%%%%%%

\begin{comment}
The AMS design at the n2EDM experiment was informed by the successfully prototyped new grid-based method, as described in Sec. \ref{sec:Method}. In this section we present the larger-scale AMS design and its technical implementation. 
\end{comment}

Given the experience with the prototype and the requirements resulting from the mapping of the experimental area, the AMS system for n2EDM was designed. Compared to the prototype, the definition of the grid structure was more constrained by the needs of n2EDM and the available space, much more ampere-turns were necessary for similar field strengths, mechanical stability was more important, and much improved quality control was needed for the construction process.

%\subsection{Mapping of the area}
%%\label{sec:4.1}
%\label{sec:Mapping}
%\input{Mapping.tex}

\subsection{AMS coil design}
%\label{sec:4.2}
\label{sec:AMS-coil-design}
The design of the AMS coil system involved several steps, with iterations: 
(i) the choice of a surface and a grid on which the coil system could be constructed around the MSR, taking into account all constraints from experimental needs, equipment and area; 
(ii) finding the currents on the grid structure to create the target homogeneous and first-order gradient fields; 
(iii) 
organization of the currents in loops and coils as well as
simplification of the optimal solution by the exclusion of (simple loop) currents contributing negligibly within the specified uncertainties.

%\textbf{Grid design.}
\paragraph{Grid design.}
 Placing the grid structure as far away as possible from the surface of the MSR ensures better field homogeneity. The main limitation was the size of the experimental area. Similarly, the whole experiment, and the MSR in particular, benefits from a clean, temperature-stabilized environment. The n2EDM experiment therefore must be separated from the main experimental hall and requires a thermal enclosure. The solution was the construction of a wooden house (`thermohouse'), similar in principle to the one of the nEDM experiment~\cite{SFC2014}. The AMS was planned to be installed on the inside of the walls of the thermohouse, which could then almost fill the experimental area. This way, optimal access to the experimental equipment and the coil system was guaranteed. The power dissipation from the coils was studied and taken into account in the lay-out of the air-conditioning system of the enclosure. It was assumed and later verified that the total the dissapated heat was quite stable for all observed external field conditions, even with dynamically controlled currents of the AMS.

 The possibility to fix the AMS to the walls and the roof of the thermohouse simplified its construction, given the mechanical stability of the structure, which was designed to carry the additional, substantial weight of the coils.

\begin{figure}
\centering
\includegraphics[width=0.5\textwidth]{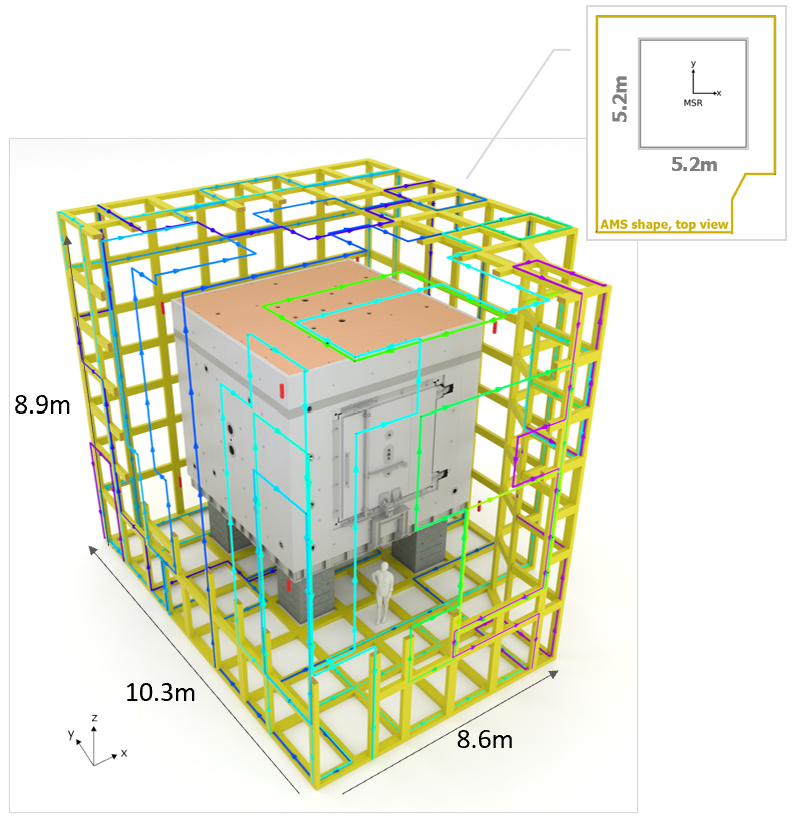}
\caption{\label{fig:AMS-shape}The AMS grid (in yellow) around the MSR. Part of the grid is not shown in the picture to allow the view onto the MSR. The colored lines represent as an example the main simple loops of the Y-coil.}
\end{figure}

Figure~\ref{fig:AMS-shape} shows the final grid of the AMS coil system with dimensions of 10.3~m $\times$ 8.6~m $\times$ 8.9~m around the MSR. It has rectangular tiles of around 1.5\,m average side-length, with a total of 308 tiles, 473 vertices, and 778 edges.

\begin{figure}
\centering
\includegraphics[width=0.45\textwidth]{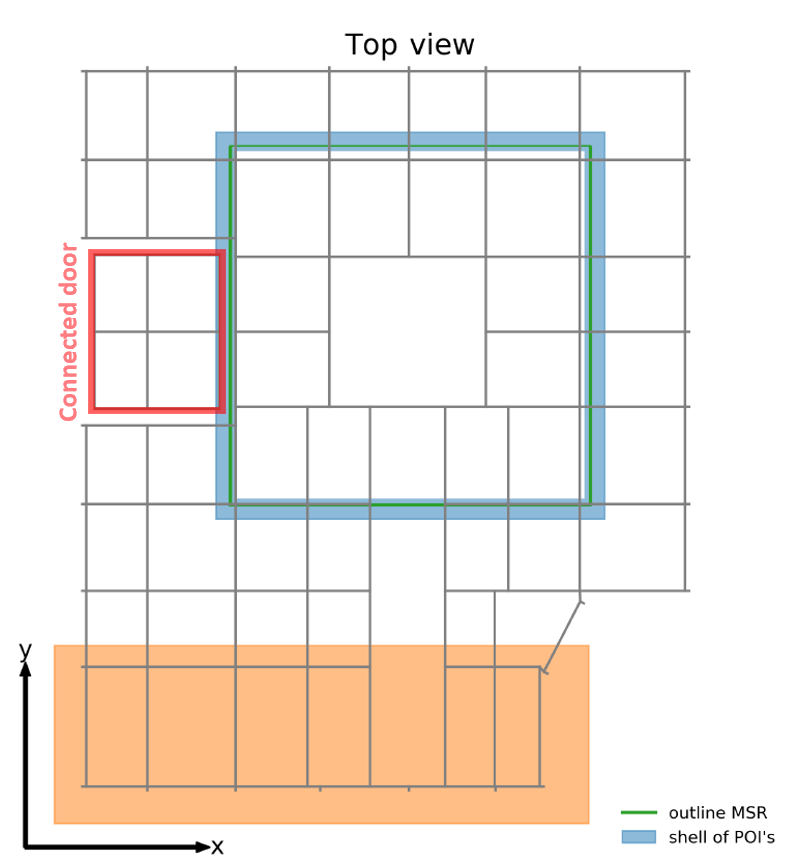}
\caption{\label{fig:VOI-Connected} Schematic top view of the AMS grid. A connected door is shown in red. The volume of the POI is indicated in blue around the MSR (green).  The orange area depicts part of the AMS, which has only a minor impact on the field homogeneity around the MSR. Adapted from~\cite{Solange2021}.}
\end{figure}

 The process of designing the grid was mainly heuristic and required several iterations.
 The density of the grid mesh was optimized~\cite{Solange2021} to achieve the best possible field-homogeneity around the MSR, while trying to keep the wir\-ing effort reasonable. To make easy access for doors and other openings for the infrastructure of the experiment, we increased the size of tiles when possible or introduced special pieces, called `connected doors' (see Fig.~\ref{fig:VOI-Connected}). A connected door is a separate grid structure placed on a detachable part, which is electrically part of a coil but, for the optimization of the circuit, topologically separated from the rest of the grid. The design method easily allows such separations.
For maintenance, it is then possible to completely detach the connected door making a larger opening to access the experiment.

Given the irregular layout of the experimental area and the thermohouse, the AMS tiles located at the kink in the wall (see Fig.~\ref{fig:AMS-shape}, lower right corner of the layout; and Fig.~\ref{fig:VOI-Connected} for the top view) would carry high currents while having little effect on the field quality. Such high currents are not ideal for the AMS system: they increase cable thickness and heating effects, requiring bulky cabling and possibly dedicated cooling.
To solve this, a regularization procedure was used in the optimization, turning the proportionality equation Eq.~\ref{Eq:basics}
into a system of equations:

%\begin{equation}
%\vec{\text{B}} = M\vec{\text{I}}
%\end{equation}

\begin{equation}
\begin{pmatrix}
\mathbf{B}\\
0
\end{pmatrix}=
\begin{pmatrix}
\mathbf{M}\\
\lambda \cdot\mathbb{1}
\end{pmatrix}
\cdot \mathbf{I}
\end{equation}

\noindent
\\
where $\mathbf{M}$ is the proportionality matrix, $\mathbf{B}$ is the target magnetic field and $\mathbf{I}$ represents the currents in the tiles. The second equation penalizes large currents. It includes the regularization parameters $\lambda$, which must be numerically determined for each coil, so a set of eight $\lambda$-values is neccessary for the full system. Optimal $\lambda$-values should then produce solutions with low total currents and still achieve the performance goal of \SI{}{\micro \tesla}-fields.

\begin{figure}
\centering
\includegraphics[width=0.45\textwidth]{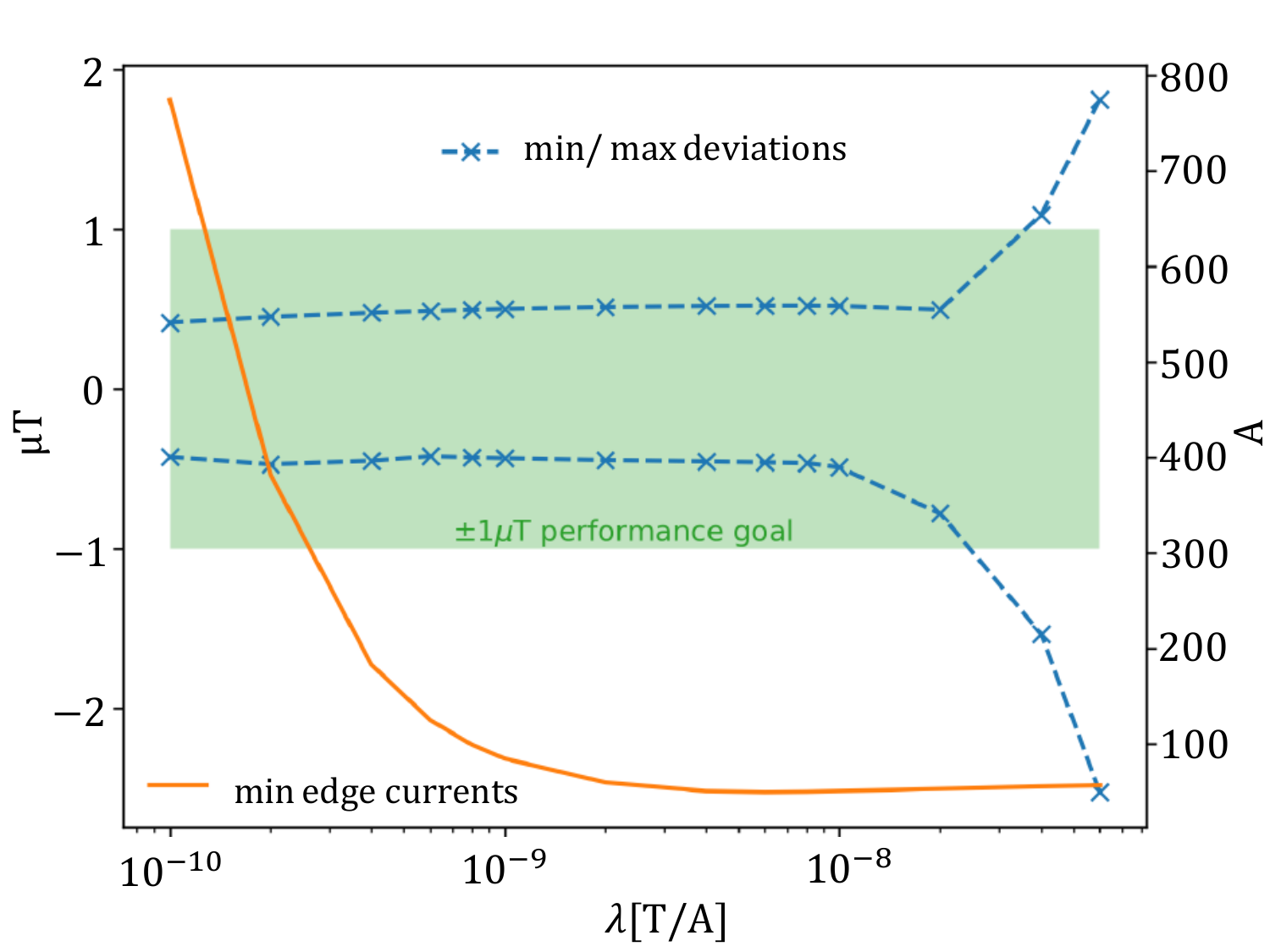}
\caption{\label{fig:lambda}
The largest deviations (`min' for negative and `max' for positive) from the target field at the MSR are plotted against the $\lambda$-parameter for the procedure of optimizing currents of the X-coil. More details in~\cite{Solange2021}. A \SI{\pm 1}{\micro \tesla} performance goal is indicated by the green box.  The axis on the right gives the average currents in the edges of the grid and the orange curve shows its decrease with increasing $\lambda$.}
\end{figure}

An example of the $\lambda$ optimization for the X-coil is shown in Fig.~\ref{fig:lambda}.
Increasing $\lambda$ reduces the mean edge-currents and thus the total current, as expected, while $\lambda$-values above $10^{-8}$ T/A cause the residual from the target field to diverge. The total current is minimal and constant for $\lambda \leq 6\cdot 10^{-9}$ T/A. Appropriate values for $\lambda$ were determined for all eight coils.

\paragraph{Choice of optimization volume.}
%\textbf{Choice of optimisation volume.}
A shell of 20\,cm thickness around the outer layer of the MSR was chosen as the volume of interest for the optimization (shown in blue on Fig.~\ref{fig:VOI-Connected}). In the real system, magnetic-field sensors are installed inside of this shell. The sensors need to be placed at a distance from the mu-metal to work reliably and the target fields need only to be reached close to the MSR surface. Making the volume for the target fields larger would require larger currents and smaller grid spacing for the AMS. Within the shell, the AMS field was numerically evaluated on random POIs, drawn from a uniform distribution. A set of 9600 POIs was used in the performance evaluation. It was checked that 
the sampling of the MSR surface by these points was sufficiently dense.
%these points sample the MSR surface densely enough.

%%%%%%%%%%%%%%%%%%%%%%%%%%%%%%%
%%%%%%%%%%%%%%%%%%%%%%%%%%%%%%%

\paragraph{Final adjustments.}
%\textbf{Final adjustments.}
The application of our method of coil design (Sec.~\ref{sec:Method}) yields a large number of simple loops on the predefined grid. These loops are not all equally important for the performance of the AMS. For the practical implementation, the number of loops can be reduced as long as the target fields can be achieved. The initial set of simple coils was ordered by importance with respect to the field intensity they produced in the volume of interest.  
The performance of the system was then recalculated for configurations where subsequently the least important simple loops were left out. 

\begin{figure}
\centering
\includegraphics[width=0.45\textwidth]{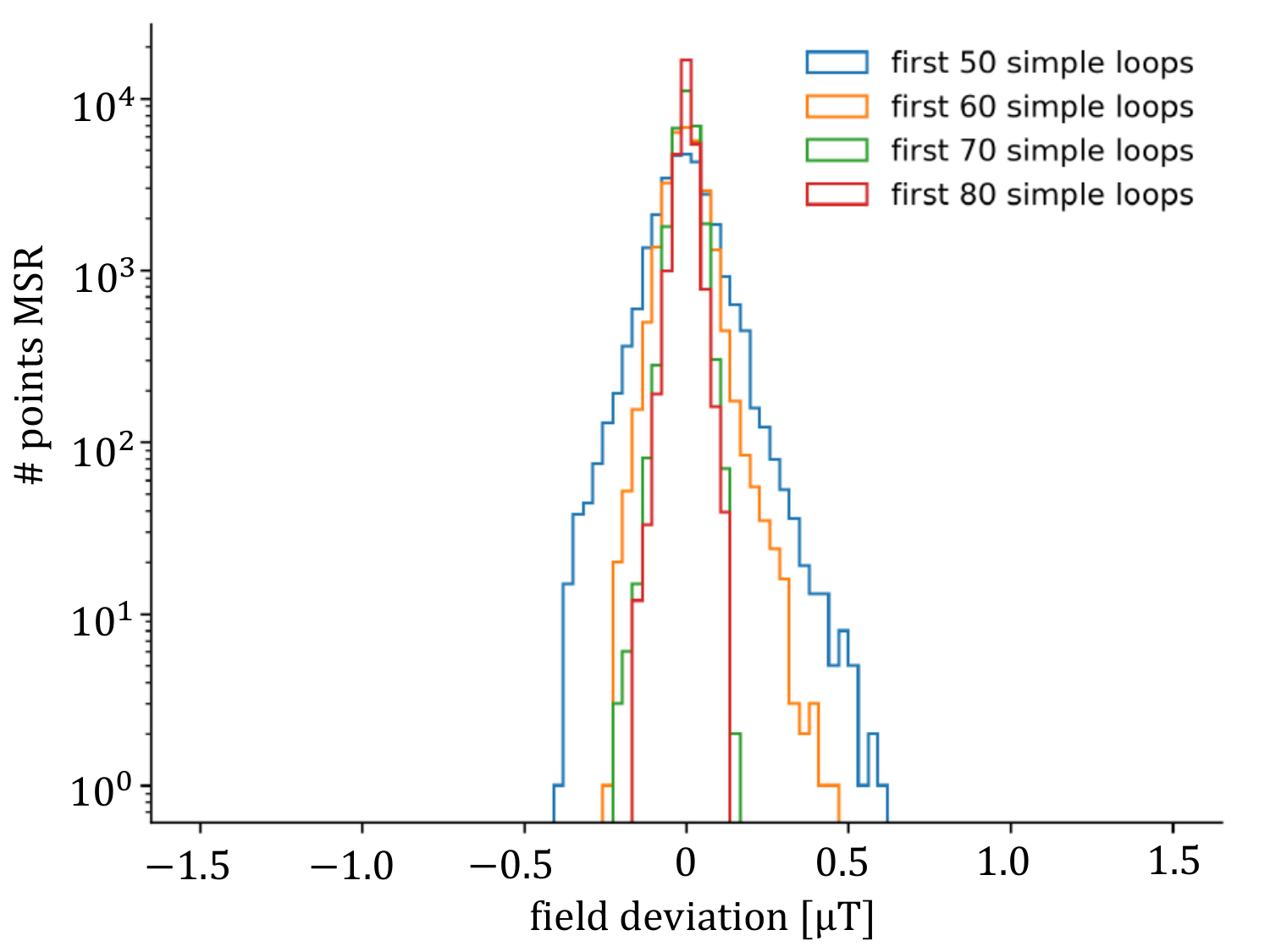}
\caption{\label{fig:Loops} Histogram of the residual fields around the MSR for different numbers of most important simple loops used in the 4th gradient coil of the system. Adapted from~\cite{Solange2021}.}
\end{figure}

As an example, Fig.~\ref{fig:Loops} shows a comparison of the residual fields at the POI close to the MSR for different numbers of the most important loops of the 4th gradient coil. The target field for the gradients is \SI{5}{\micro\tesla / \meter} and residuals refer to the `full field' configuration producing this gradient. The best solution for this coil had 123 simple loops. However, the performance stays close to optimal down to a reduced number of the 70 most important loops.
This procedure was used for all eight AMS coils, yielding a reduction of 40\% in the total number of simple loops while still reaching the target fields.

\paragraph{Complete system.}
%\textbf{Complete system.}
The final design of the AMS system comprises eight coils matched to the optimized grid. Each of the coils consists of 50-70 simple loops 
%which could be operated in series in order 
to achieve the target fields. An example of a subset of the main simple loops of the Y-coil is shown in Fig.~\ref{fig:AMS-shape}. 
As in the prototype, each coil consists of three electrical circuits with large, medium, and small elementary currents. This minimized wiring effort, total weight, and power dissipation, while keeping the number of current sources at a manageable level.
Table~\ref{tab:Coils} summarizes the chosen elementary currents and the numbers of simple loops for all coils.

\begin{table}
	\renewcommand{\baselinestretch}{1.5}
	\renewcommand{\arraystretch}{2}
	\centering
 \begin{tabular}{|c|c|c|}
\hline	
Coil & Currents $I$ &  $N$ loops \\
\hline
X & 15A / 5A / 1A  & 59 \\

Y & 15A /5A/1A & 68 \\

Z & 15A / 5A / 1A & 60 \\

1G & 10A / 3A / 1A & 50 \\

2G & 12A / 5A /1A & 69 \\

3G & 15A / 4A /1A & 60 \\

4G & 8A / 3A /1A & 70 \\

5G & 12A / 5A / 1A & 69 \\
\hline
\end{tabular}
\caption{\label{tab:Coils} \linespread{1}\selectfont Values of the three elementary currents $I$ and numbers of simple loops $N$ for the eight coils of the final AMS design.}
\end{table}

\begin{figure*}
\centering
\includegraphics[width=0.8\textwidth]{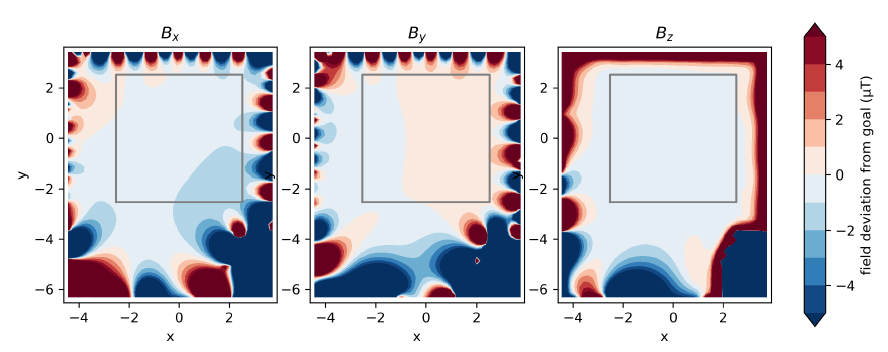}
\caption{
\label{fig:simulation}
Two-dimensional maps of the calculated magnetic-field residuals at z = 0, which is the vertical center of the MSR. The x and y coordinates are given in meters. The outline of the MSR is depicted as a gray square. The residuals are plotted for the full field in all coils including cross compensation, see text. Adapted from~\cite{Solange2021}.}
\end{figure*}

The calculated residual fields from the target field of the finalized coils are shown in Fig.~\ref{fig:simulation} for the so-called `full-field' configuration, in which all coils are powered to simultaneously produce the homogeneous fields of \SI{50}{\micro\tesla} in each spatial direction as well as the five \SI{5}{\micro\tesla / \meter} gradients, see Tab.~\ref{tab:AMS-Cartesian-harmonic-polynomials}. Each coil was designed individually to compensate one of these eight basis fields. Due to the discretization and the simplification inherent to the design method, the fields generated by the coils slightly deviate from their target fields, within the allowed ranges. Such deviations will add up (vectorially) when operating the coils together. Large deviations do not usually occur at the same place and in the same directions, they appear rather more randomly, leading even to some `cross compensation', with the result displayed  in Fig.~\ref{fig:simulation}. The residual of each coil's field from its target value can itself be represented as an expansion in the same basis fields produced by the other coils (plus neglected higher-order fields). Therefore, when measuring the response matrix for the built system, these effects are taken into account automatically.

\hspace{-1ex}Other, potentially important aspects of the real-world AMS system are the unavoidable imperfections of the current paths. One of these issues arises from the bending of cable bundles at each vertex. Obviously, these are not 90 degree corners but require bending radius up to 10\,cm. Another imperfection is the position of the wires in the bundles with respect to the ideal grid. On some edges of the system, the area crossed by all cables was up to 80\,cm$^2$; necessarily some wires are displaced by several cm from their ideal position. All these effects were simulated and in all cases we concluded that they were tolerable or even negligible. Again, aforementioned cross-compensation helps considerably: while a coil might slightly deviate from its ideal performance, it will still be completely linearly independent of the other seven coils and the system can function almost equally well.

\subsection{The AMS technical implementation}
\begin{figure}[ht]
\centering
\includegraphics[width=0.45\textwidth]{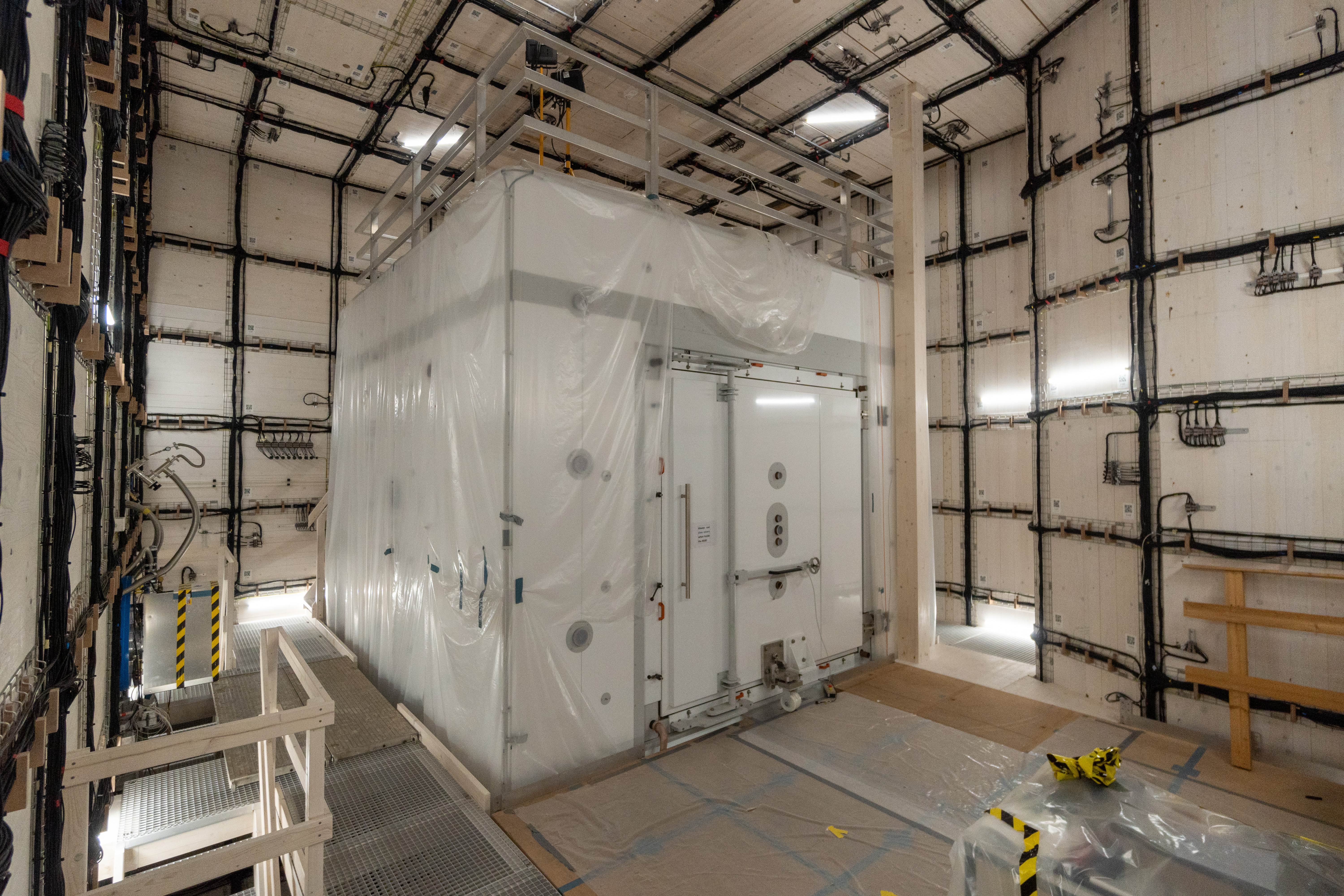}
\caption{\label{fig:AMS-built} Photograph of the AMS coil system constructed on the walls of the thermohouse around the MSR. On the right-hand side, in the middle tiles, and on the back-side to the left of the MSR, a few DIN rails are visible which are wired to connect circuits, as explained in the main text. Of course, the AMS extends fully around the MSR, also on the entire floor. The  platform visible in front of the MSR is about 2.5\,m above floor level.}
\end{figure}

%\label{sec:4.3}
\label{sec:Technical-implementation}
The AMS coil system was mounted in the thermohouse of n2EDM over the course of approximately one year. An over\-view  of the construction process is presented here, while more details are found in Ref.~\cite{Solange2021}.
Figure~\ref{fig:AMS-built} shows a picture of the finalized system. 

The AMS grid structure was formed by cable trays out of stainless steel, mounted directly onto the inner walls of the wooden thermohouse around the MSR. The cable trays were grounded in a way to inhibit closed loops and eddy currents through the trays.
In order to mount
the coils of Table~\ref{tab:Coils} efficiently onto the grid (Fig.~\ref{fig:AMS-shape}) along their calculated paths, one could not simply wind long cables. Instead, cables of a simple loop were bundled and installed on the cable trays along the roughly 500 different paths. The ends of the wires of the bundles were carefully crimped to form the loops. All the simple loops for each elementary current were connected in series such that a `coil' consisted of three independent circuits.
The installation of a bundle carefully followed a detailed plan, connecting it at some start position in the thermohouse and following a prescribed path along numbered vertices. The start and the end position of the circuits were later connected to terminals on DIN rails, which themselves were connected appropriately with interconnection wires.   At each vertex of the grid the correct direction had to be checked, going straight or around a corner. 
A system of bar-codes on the wires and QR-codes near the vertices on the walls, both completed with human-readable names, along with a dedicated smartphone scanning-app, were developed for a continuous verification and quality control during the installation. Completed circuits were electrically checked and DC resistances were measured to guarantee quality of crimp connections.
This way, a total of \SI{55}{\kilo \meter} of wire was installed, without any indications of error. 

\subsection{AMS current sources}
\label{sec:Amplifiers}
To power the AMS coils, we have designed and built bipolar high-power current sources in-house at PSI, based on APEX-PA93 linear operational amplifiers~\cite{APEX}. 

\begin{figure}
\centering
\includegraphics[width=0.45\textwidth]{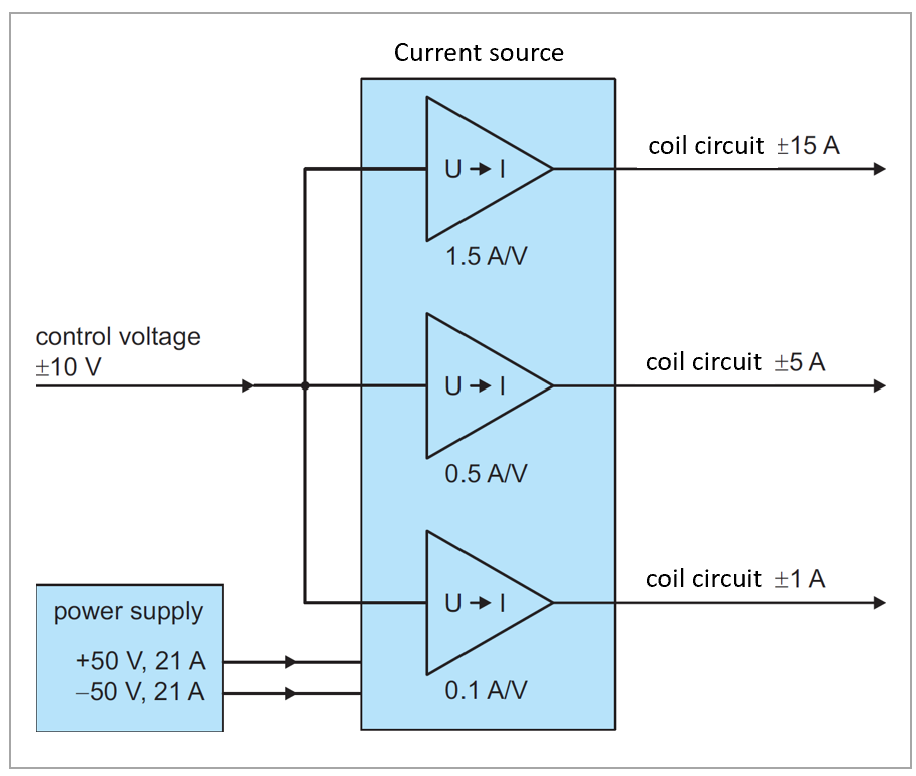}
\caption{\label{fig:amplifier} Simplified scheme of the bipolar current source, developed to power AMS coil. Each current source consists of three channels (here: 15A, 5A and 1A), with their output currents proportionally to their individual control voltage.}
\end{figure}

Each current source consists of three channels, delivering the elementary currents to the corresponding three circuits of a coil (Fig.~\ref{fig:amplifier}). The currents of all three channels change proportionally to their control voltages, which can be set in the range from \SI{-10}{\volt} to \SI{10}{\volt}. This allows for an efficient realization of the three-fold powering approach described in Sec.~\ref{sec:Method}.
For coils with different design currents (see Table~\ref{tab:Coils}) the software will command them with properly reduced values.
Depending on the channel, up to six APEX amplifiers were connected in parallel and complemented by a system of matched resistors to deliver the required output current and distribute the power dissipation.
An internal stabilization network combined with external damping resistors enables the current source to drive inductive loads up to ~\SI{1}{\henry}.
This is important because although the self-inductance of the coils ranges only from \SI{3}{\milli \henry} to \SI{75}{\milli \henry}, their mutual inductance is up to \SI{500}{\milli \henry}.

Each of the current sources is supplied with \SI{\pm 50}{\volt} from an external switching power supply with a large filter capacitor to ensure a low-noise level of operation. The total heat dissipation in each of the current sources can reach up to \SI{500}{\watt}, which is removed by an efficient built-in cooling system.
%, enabling optimal performance.
%

As part of the performance verification of the current sources, they were connected to the coils and their responses measured to a square-shaped input signal with the maximum amplitude of \SI{10}{\volt}.  The output signals reached their maxima with typical time constants of approximately \SI{80}{\ms}, fast enough for dynamic AMS operation in the sub-Hertz frequency range, as required.

\subsection{Fluxgates sensors}
\label{sec:Fluxgates}

Eight 3-axis SENSYS fluxgate sensors~\cite{SENSYS} were installed around the MSR to measure the magnetic field and provide feedback information for the dynamic mode of the AMS.
The initial optimization of their positions was carried out similarly to the one of the prototype (Sec.~\ref{sec:Prototype-Performance}) and positions close to the corners of the MSR were found. 

In Sec.~\ref{sec:AMS-shielding}, results of the dynamic shielding performance are reported, based on these positions of the fluxgates and the control system, described in the next section. 

\subsection{Control system}
\label{sec:ControlSystem}
% !TeX spellcheck = en_GB

The control system is based on Beckhoff modules ELM3148 (24 bit ADCs) and EL4134 (16 bit DACs) operating at 1kHz.
The readings from all fluxgate channels are stored in the array $\boldsymbol{B}_{\mathbf{m}}$ that has up to 51 entries.
There is a minimal delay of two cycles for any reaction. Thus, the currents $\boldsymbol{I}$ of the next cycle $[i+1]$ are calculated as:
\begin{equation}
	\boldsymbol{I}[n+1] = \boldsymbol{I}[n] +  \boldsymbol{k} \times  ( \boldsymbol{M}^{-1} \times (  \boldsymbol{B}_{\mathbf{t}} - \boldsymbol{B}_{\mathbf{m}}[n-1] )), 
\end{equation}
%
%\begin{equation}
%	\boldsymbol{I}[n+1] = \boldsymbol{I}[n] +  \boldsymbol{k} \times  ( \boldsymbol{M}^{-1} \times (  \boldsymbol{B}_{\mathbf{T}} - \boldsymbol{B}_{\mathbf{M}}[n-1] )), 
%\end{equation}
%
where $\boldsymbol{B}_{\mathbf{t}}$ is the target field, normally chosen as 0.
The feedback matrix $\boldsymbol{M}^{-1}$ is the pseudo-inverse (calculated offline as explained in  Eq.\,(\ref{Eq:Inverse})) of the response matrix $\boldsymbol{M}$. 
The latter is obtained by scanning all coil currents individually and analysing their response by linear regression (as described in Sec.~\ref{sec:Prototype-Performance}).
A multiplication constant $\boldsymbol{k}$ slows down the feedback to avoid oscillations. We use the same value of 0.013, found empirically, for all eight coils.
This results in a characteristic time constant of about \SI{50}{\ms}. 
A faster operation is prevented by the current sources, but was never intended.
%
%By reducing the $k$ temporarily by a factor 100, the AMS response becomes slow enough that it does not react on the degaussing signal. 
%
%Thus for the first time, keeping the AMS running (though in slower mode) during degaussing became  possible.

While the performance is 
satisfactory already, potential improvements will be studied once commissioning of other n2EDM subsystems, which it would interfere with, is completed. 
A more detailed simulation model with improved utility for various numerical studies is being deployed, additional fluxgate sensors are being installed, and further optimization of sensor positions and feedback algorithms pursued.
%

\begin{comment}
, see Table~\ref{tab:fgpositions}.

%Almost all the fluxgate positions lie close to the corners of the MSR (see Table~\ref{tab:fgpositions}), which was optimized similar to the AMS prototype (Sec.~\ref{sec:Prototype-intro}) and turned out to be sufficient for stable dynamic performance of the AMS. 

\begin{table}[htb]
	\renewcommand{\arraystretch}{2}
    \centering
    \caption{Positions of the AMS feedback fluxgates in the n2EDM coordinate system (see Fig.~\ref{fig:AMS-shape}). The MSR center is at (0,0,0), {\bf the coordinates of its corners are (....).}}
    \vspace{\baselineskip}
    \begin{tabular}{|l|l|l|l|l|}
        \hline
         & x [m] & y [m] & z [m] \\
                 \hline
         FG1 & -0.36 & -2.66 & 0.06\\
         FG2 & -2.70 & 2.70 & -2.10\\
         FG3 & 2.70 & -2.10 & -1.90\\
         FG4 & 2.00 & 2.20 & -2.45\\
         FG5 & -2.70 & -2.70 & 2.70 \\
         FG6 & -1.50 & 2.70 & 2.70\\
         FG7 & 2.70 & -1.80 & 2.70\\
         FG8 & 2.10 & 2.70 & 2.70\\ \hline
    \end{tabular}
    \label{tab:fgpositions}
\end{table}

\begin{figure}[ht]
\centering
\includegraphics[width=0.45\textwidth]{Images/Feedbacmatrix-AMS.png}
\caption{\label{fig:matrix-AMS} Visualisation of the AMS feedback matrix}
\end{figure}

\end{comment}

\section{The AMS performance measurements}
%\label{sec:5}
\label{sec:AMS-performace-intro}
After the installation and the commissioning of all AMS coils and power supplies, we validated the static magnetic-field generation and measured the shielding performance of the AMS, as described below.

\subsection{Magnetic fields generated by the AMS coils}
%\subsection{Validation of the AMS coils}
%\label{sec:5.1}
\label{sec:Validation}
As the MSR was installed in the experimental area before the AMS system, the actual magnetic fields generated by the AMS coils are not the simple homogeneous and first-order gradient fields as designed, but are modified by the MSR. 

Thus, after the quality control described in Sec.~\ref{sec:Technical-implementation}, which guaranteed the proper pathways for the currents, actual mag\-net\-ic-field measurements were compared to results of a FEM simulation model implemented in COMSOL~\cite{COMSOL}. We used the design fields as imported background fields and the outermost mu-metal surface of the MSR as a 10\,cm thick  layer of high magnetic permeability. It was verified that above a certain permeability and thickness, the results of the simulation became independent of these details. 

As the experimental area around the AMS was already partially occupied by other equipment after its completion, magnetic-field measurements around the MSR had to be done in a sampling mode rather than in form of full field maps (as described for the empty area in Sec.~\ref{sec:Mapping}). 

\begin{figure}[t]
\centering
    \includegraphics[width=0.45\textwidth]{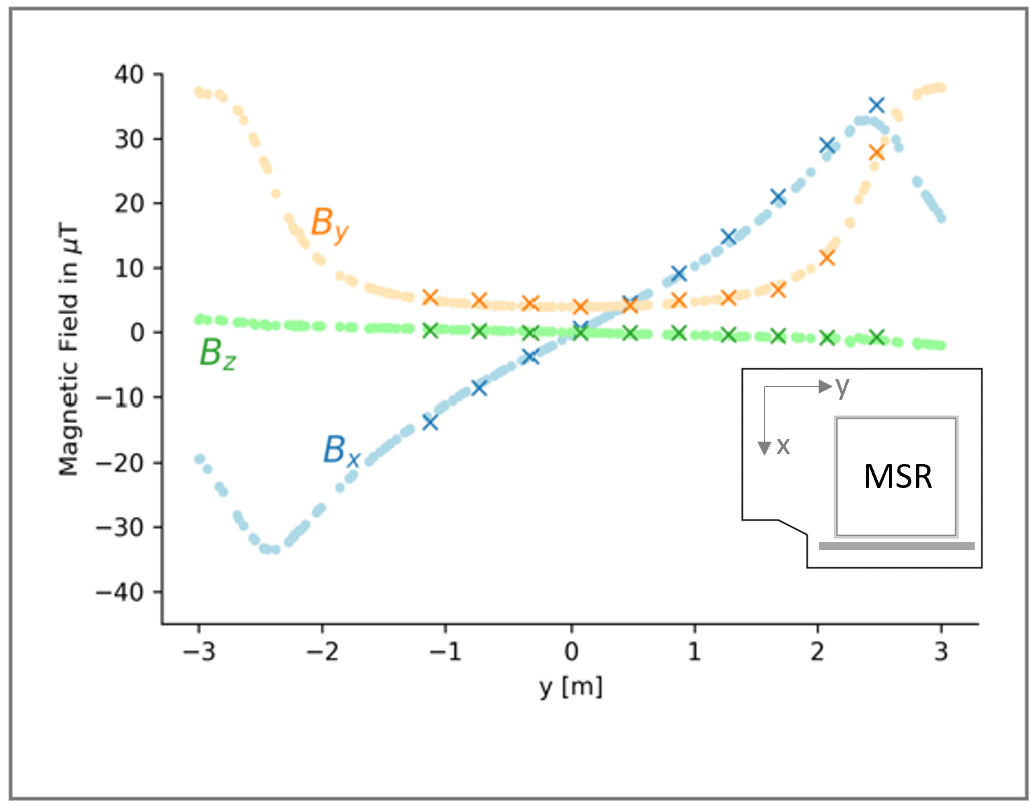}
  \caption{Example of an AMS validation measurement: comparison between the measured (crosses) and the simulated (dotted line) magnetic field values for $B_x$, $B_y$, and $B_z$ components produced by the Y-coil. The measurements were taken along the gray line in the inset ($y$-direction) at $z=$\SI{-1}{\metre}. Adapted from~\cite{Solange2021}.}
\label{fig:AMS-mapping}
\end{figure}

The magnetic fields created by individual coils were measured in some selected, easily accessible areas, usually along a straight aluminum profile with one fluxgate, and compared to the simulation.  Figure\,\ref{fig:AMS-mapping} shows an example of such a comparison. The measured and the simulated magnetic-field values for the $B_x$, $B_y$ and $B_z$ components produced by the Y-coil are shown. The current of the Y-coil was pulsed on and off for the measurement to enable proper background-field subtraction. The coil current was chosen to be half of the maximum current. The measurements were taken along the grey line shown in the inset, at a height of $z=$\SI{-1}{\metre} (below the center of the MSR) in the y-direction at a distance of about 20\,cm from the MSR surface. 
%The  simulations were performed in COMSOL~\cite{COMSOL} and took into account effect of the mu-metal.

\begin{comment}
\begin{figure*}
\centering
    \includegraphics[width=0.7\textwidth]{Images/n2edm-map.png}
  \caption{Example of the AMS validation map: comparison between the measured (crosses) and simulated (dotted line) magnetic field values for $B_x$, $B_y$ and $B_z$ components produced by the y-coil. The measurement was taken along a grey line in y-direction relative to the MSR. More details in [PhD Solange].}
\label{fig:AMS-mapping}
\end{figure*}
\end{comment}

The result of the measurement agrees with the simulations, and the behaviour of the magnetic-field components is as expected for this example. The design field at the sampled positions without MSR would only have a $B_y$ component, with $B_x=B_z=0$. One can see this feature emerging for large positive and negative values of $y$. While the non-existent $B_z$ component is unaffected by the mu-metal shield, the $B_y$ component gets absorbed into the mu-metal by drawing it into the $B_x$ component with maximal amplitude at the edge of the MSR around $y\approx \SI{-2.8}{m}$. A similar qualitative and quantitative agreement was observed for the other coils, which confirms a good understanding of the AMS coil system as built.

\subsection{AMS shielding measurements}
%\label{sec:5.2}
\label{sec:AMS-shielding}
Commissioning measurements of the dynamic AMS shielding were performed outside and inside the MSR during ramps of the superconducting high magnetic-field facility `SULTAN'~\cite{SULTAN1981}. Magnets of this facility were already of concern to the predecessor nEDM experiment~\cite{SFC2014}. The facility is about 30\,m away from the n2EDM experiment. Its magnets can ramp up to $\SI{11.5}{\tesla}$, producing fields up to $\SI{40}{\micro\tesla}$ at the MSR front and back walls mainly in horizontal direction with the AMS system off.

\begin{figure*}[t]
	\centering
	\includegraphics[width=0.8\textwidth]{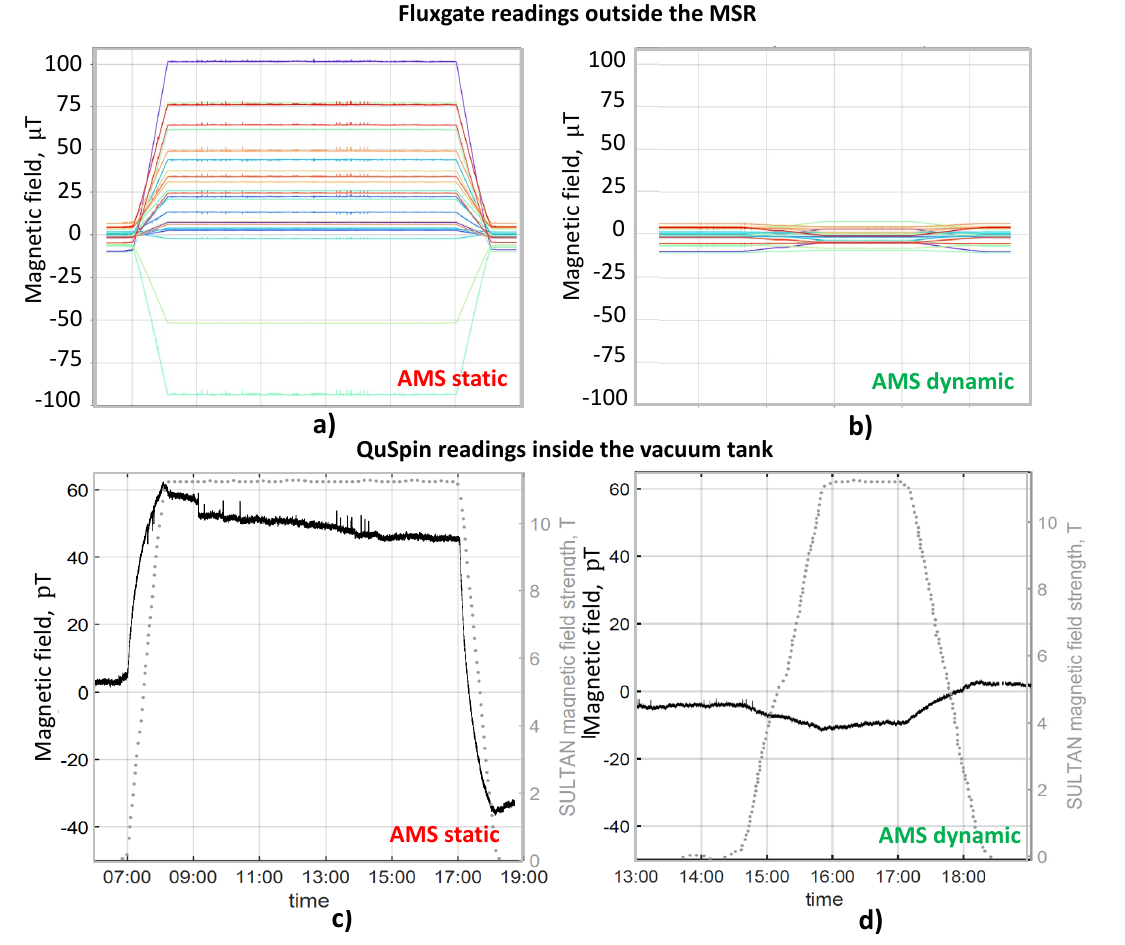}
	\caption{The AMS suppression of magnetic fields from SULTAN: (a-b) - magnetic fields measured by
		the feedback fluxgates outside the MSR during two different SULTAN ramps with the AMS system in static (a) and
		dynamic (b) modes; (c-d) - 
		%{\bf CHECK (c): looks like the two curves are not on exactly the same time scale - QuSpin starts to go down before the SULTAN ramp ??? SULTAN ramp should start at 16:00? That would also fit (a). How was the synchronization done??} 
		the magnetic fields of the SULTAN magnet for the two ramps (dotted grey line, right scale) along with the corresponding magnetic field measured by an optically-pumped (QuSpin) magnetometer~\cite{QuSpin} inside the MSR (black line, left scale). %{\bf perhaps try to do a properly scaled plot of the ratio of the ramps in (c) to get shielding factor over external field strength? ... the curvature of the QuSpin ramps is strange: most likely it is not the changing shielding factor but the drifting field as otherwise both ramps should be convex ... ? Or perhaps: both times the field change starts fast and then slows down because both times the MSR is kind of in equilibrium ...?}
	}
	\label{fig:SULTAN-all}
\end{figure*}
  %This is already included in Validation.tex  to provide a better placeing of the floating figure

The magnetic fields outside the MSR were measured by the eight 3-axis SENSYS fluxgates~\cite{SENSYS} involved in the feedback, as previously described, and by several additional monitor fluxgates. %The latter were placed in front of the center of the xz door-side of the MSR, see Figs.~\ref{fig:AMS-shape} and~\ref{fig:AMS-built}, the closest MSR surface in direction of SULTAN. 
The magnetic field inside the MSR was measured by a much more sensitive optically-pumped QuSpin magnetometer (Gen3, zero-field configuration)~\cite{QuSpin}. It was placed roughly at the center of the MSR with one of its two sensitive directions along the $z$-axis, the most relevant for nEDM measurements, and the other one was oriented along the direction of the largest SULTAN perturbation. 
A second QuSpin sensor, installed close to the first one, was used to ensure that field changes could be identified as such and readings of one sensor were not simply due to sensor drift.

Figure~\ref{fig:SULTAN-all} shows magnetic-field values measured during the SULTAN ramps with the AMS in static mode
and dynamic mode, respectively. 
In static mode, Fig.~\ref{fig:SULTAN-all} (a), the background field was compensated only approximately some time before the ramp, keeping AMS currents constant.
Thus, the initial spread of the fluxgate readings was not illustrating an optimal zero-field setting.
During the SULTAN ramp, the measured magnetic field changed from several \SI{}{\micro \tesla} up to roughly 100\,\SI{}{\micro \tesla}, depending on the positions of the fluxgates. The fluxgates positioned near the corners of the MSR, fields get amplified and are at some points much larger than in the empty-area mapping (Sec.~\ref{sec:Mapping}).
When the AMS system is operated in dynamic mode, Fig.~\ref{fig:SULTAN-all} (b), the corresponding magnetic-field changes in the feedback fluxgates are reduced to a level of a few \SI{}{\micro \tesla}.

The measurements with the QuSpin sensor, shown in Fig.~\ref{fig:SULTAN-all} (c), additionally show the passive shielding of the MSR. As determined with the earlier mapping campaign, the field variation from the SULTAN ramp at the location of the QuSpin sensor would be about \SI{40}{\micro \tesla} without the MSR. A rough analysis of the magnetic field as measured by the QuSpin inside the MSR, during the ramps, found field changes of about 60 -- 80\,\SI{}{\pico \tesla}. This would correspond to a quasi-static shielding factor of the MSR of roughly $(5-7) \times 10^5$. It is well known that the magnetic-shielding performance of such magnetic shields improves for larger field variations due to the increase of permeability $\mu_{\mathrm{r}}$ for larger magnetic-field strength $H$ until saturation effects set in. It is therefore very important to describe the excitation field when quoting a shielding factor. For this particular example, the MSR of n2EDM has a quasi-static shielding factor of $1 \times 10^5$ for an excitation field corresponding to \SI{\pm 2}{\micro \tesla} at the unshielded sensor location~\cite{MSR2022}. This is the relevant shielding factor for n2EDM, as \SI{}{\micro \tesla}-size perturbations will still be possible, even with a perfectly functioning AMS.

When the AMS system is in dynamic mode, the Qu\-Spin sensor measures a more attenuated signal, see Fig.~\ref{fig:SULTAN-all} (d). The amplitude of this remaining field change is about \SI{8}{\pico \tesla}, a factor of 7--10 smaller, compared to the SULTAN ramp when the AMS is in static mode. It is, however, not straightforward to take this as the shielding factor of the AMS system alone, as we just saw that the passive, quasi-static shielding performance of the MSR depends on the size of field perturbations on the shield, which in turn depends on AMS performance. 

Nevertheless, we can deduce the approximate shielding factor for the combined system of the AMS and the MSR for large and slowly changing perturbations (here a one-hour ramp to  \SI{40}{\micro \tesla}) as
$5 \times 10^6$. More importantly, the result demonstrates that the goal of suppressing field changes down to below 10\,pT inside the MSR was achieved, which was set up as a requirement for the n2EDM experiment.

Another interesting observation from the comparison between
Figs.~\ref{fig:SULTAN-all} (c) and (d) is that at least in this set of measurements it appears that the larger field variation on the outside of the MSR in the static case caused the field inside of the MSR to drift more, about 30--40\,\SI{}{\pico \tesla}, compared to the dynamic case with a drift of less than~\SI{10}{\pico \tesla}. 
One can see, that the drift following the ramp-up in (c) is opposite to the induced change, while the drift following the ramp-down is in opposite direction. 
This is expected from the reaction of the mu-metal layers of the MSR to the perturbation. 
Such drifts are part of unwanted behaviour of a passive magnetic shield, which, when exposed to large external field variations, slowly absorbs the remanent field until it reaches the state of lowest energy. 
The AMS system largely reduces the impact of such effects.

\section{Summary}
\label{Summary}
The AMS system was designed and built to compensate homogeneous and first-order gradient external magnetic-field  changes around the MSR of the n2EDM experiment. It was developed using a novel method of coil design. After successful prototyping at ETH Zurich, the AMS system was constructed and commissioned at the n2EDM experiment at PSI. First performance measurements demonstrated its ability to suppress magnetic-field changes of about \SI{\pm 50}{\micro \tesla} (homogeneous) and \SI{\pm 5}{\micro\tesla / \meter} (first-order gradients), to the level of a few \SI{}{\micro \tesla}. The optimization of the AMS system using measurements and improved simulations, e.g., concerning fluxgate positioning and feedback algorithm, is ongoing and might further improve its performance. In any case, with the performance demonstrated in this paper, the combined  system of AMS and MSR meets the specifications of the n2EDM experiment, providing a magnetic-field stability within the neutron volume at the 10\,pT level.

\section{Acknowledgements}
We gratefully acknowledge the support provided by ETH and PSI technicians, electrical engineers and electricians.  In particular, we appreciate the efforts of M.~Meier, L.~Noorda, A.~Angerer, R.~Wagner, S.~Hug, R.~Schwarz,
M.~Ettenreich, L.~Künzi, D.~Di~Calafiori, P.~Bryan, E.~Hüsler, R.~Käch, H.~Scheppus.

We acknowledge financial support from the Swiss National Science Foundation through projects No. 117696 (PSI),
No.~137664 (PSI), No. 144473 (PSI), 
No.~157079 (PSI), No. 172626 (PSI), No.~126562 (PSI), No.~169596 (PSI), No. 178951 (PSI), No.~181996 (Bern), No.~162574 (ETH), No.~172639 (ETH), No.~200441 (ETH). 
The group from Jagiellonian
University Cracow acknowledges the support of the National Science Center, Poland,
Grants No.~UMO-2015/18/M/ST2/00056, No.~UMO-2020/37/B/ST2/02349, 
\\ and
No.~2018/30/M/ST2/00319, as well as by the Excellence Initiative – Research University Program at the Jagiellonian University. This work was supported by the Research Foundation-Flanders (BE) under Grant No. G.0D04.21N. Collaborators at the University of Sussex acknowledge support from the School of Mathematical and Physical Sciences, as well as from the STFC under grant ST/S000798/1.We acknowledge the support from the DFG
(DE) on PTB core facility center of ultra-low magnetic field KO 5321/3-1 and TR 408/11-1.
We acknowledge funding provided by the Institute of Physics Belgrade through a grant by the Ministry of Education, Science and Technological Development of the Republic of Serbia. This work is also supported by Sigma Xi grants $\#$ G2017100190747806 and
$\#$ G2019100190747806, and by the award of the Swiss Government Excellence Scholarships (SERI-FCS) $\#$ 2015.0594.

%\clearpage

\bibliographystyle{spphys}

\def\urlprefix{}
\newcommand{\doi}[1]{doi: \href{https://doi.org/#1}{\nolinkurl{#1}}}
%\printbibliography
%\bibliography{Ref-short}
\bibliography{Zotero}
%\bibliographystyle{alpha}
%\bibliography{sample}

\end{document}